\documentclass[onecolumn,nofootinbib,superscriptaddress,eqsecnum,showpacs,prd, floatfix,preprintnumbers,amsmath,amssym,12pt]{revtex4}

\pdfoutput=1
\usepackage {natbib}
\usepackage {graphicx}
\usepackage {color}
\usepackage{bm}

\usepackage{amsmath, amsthm, amsfonts}
\usepackage{hyperref}

\begin{document}

\title{The stochastic background from cosmic (super)strings: \\popcorn and (Gaussian) continuous regimes}

\author{Tania Regimbau}
\email{regimbau@oca.eu}
\affiliation{ARTEMIS, Observatoire de la C\^{o}te d'Azur, Universit\'{e} de Nice Sophia-Antipolis, CNRS, 06304 Nice, France}
\author{Stefanos Giampanis}
\email{gstef@gravity.phys.uwm.edu}
\affiliation{Center for Gravitation and Cosmology, Department of Physics,
University of Wisconsin-Milwaukee, P.O. Box 413, Wisconsin, 53201, USA}
\author{Xavier Siemens}
\affiliation{Center for Gravitation and Cosmology, Department of Physics,
University of Wisconsin-Milwaukee, P.O. Box 413, Wisconsin, 53201, USA}
\author{Vuk Mandic}
\affiliation{School of Physics and Astronomy, University of Minnesota-Twin Cities, USA}
\date{\today}

\begin{abstract} 
In the era of the next generation of gravitational wave experiments a stochastic 
background from cusps of cosmic (super)strings 
is expected to be probed and, if not detected, to be significantly constrained. 
A popcorn-like background can be, for part of the parameter space, 
as pronounced as the (Gaussian) continuous  contribution from unresolved sources that overlap in frequency and time. We study both contributions 
from unresolved cosmic string cusps over a range of frequencies relevant to 
ground based interferometers, such as LIGO/Virgo second generation (AdLV) and 
Einstein Telescope (ET) third generation detectors, the space antenna LISA and 
Pulsar Timing Arrays (PTA). We compute the sensitivity (at $2 \sigma$ level) in the parameter space for AdLV, ET, LISA and PTA. We conclude 
that the popcorn regime is complementary 
to the continuous background. Its detection could therefore enhance confidence 
in a stochastic background detection and possibly help 
determine fundamental string parameters such as the string 
tension and the reconnection probability.
\end{abstract}

\pacs{11.27.+d, 98.80.Cq, 11.25.-w, 04.80.Nn, 04.30.Db, 95.55.Ym, 07.05.Kf}

\maketitle

\section{Introduction}

Cosmic (super)strings, formed as linear topological defects during symmetry breaking 
phase transitions~\cite{alexbook,hin95}, or in string theory inspired inflation 
scenarios~\cite{jon02,jon03,sar02,dva04,cop04}, may produce strong bursts of gravitational waves (GW)~\cite{dam00,dam01,dam05,sie06,sie07}. 
In particular, the emission from cusps, where, for a short period of time, the
string reaches a speed very close to the speed of light, in oscillating 
cosmic (super)string loops, may be strong enough to be detected by the next generation 
of ground based detectors such as advanced LIGO/Virgo (AdLV) \cite{har10,AdLIGO,AdVIRGO} 
and the planned Einstein Telescope (ET)  \cite{ET}, the space antenna LISA \cite{LISA} 
or Pulsar Timing Arrays (PTA) such as PPTA in Parkes/Australia, NANOGrav in North America, or EPTA in Europe \cite{PTA}. 

If strings can inter-commute and form loops that decay
gravitationally or through some other channel (for example, Abelian strings), the network evolves
toward a scaling regime \cite{1998PhRvD..57.3317M,1997ApJ...491L..67S}, in which 
the statistical quantities that describe the network, such as the typical distance 
between strings and the average size of loops produced by the network, scale with the cosmic time, 
and the string energy density is a small constant fraction of the radiation or matter density. 
This regime is possible because the network produces loops which decay by radiating gravitationally, and take
energy out of the network. 
There remains some uncertainty in the size of loops produced by a cosmic string network. One 
possibility is that the size of loops is set by the gravitational back-reaction scale, the scale of 
perturbations on long cosmic strings~\cite{Dubath:2007mf,Siemens:2002dj}. In this case the loops 
produced are sufficiently small that they radiate away radiate away the energy associated with their length in less than, or of order, 
a Hubble time. Another possibility, suggested by recent numerical 
simulations~\cite{BlancoPillado:2011dq},
is that loops form at a much larger size comparable to the Hubble length at the time of formation. In this case 
loops live for a long time decaying gravitationally in many Hubble times.

In this work we treat the strings as one-dimensional objects (zero width, or Nambu-Goto approximation) and we will consider the case of small loops, leaving the large loop case for a future publication. The rate and the amplitude 
of the bursts depend on three parameters, the string tension $\mu$ (we consider Nambo-Goto strings), the reconnection 
probability $p$ and the typical size of the closed loops produced in the string 
network $\varepsilon$. 
The closest sources can be detected individually, while unresolved sources at higher redshift  
contribute to a stochastic background, with a popcorn-like noise on top of a (Gaussian) continuous background~\cite{dam00,dam01,dam05,sie06,sie07}. In this paper, we investigate the GW signal for a grid of values in the parameter space. We compare the popcorn and the continuous contributions over a range of frequencies from $10^{-12}-10^4$ Hz and discuss the constraints that could be placed on the parameters by AdLV, ET, LISA and PTA, using the standard cross correlation statistics \cite{all99}. 

For cosmic (super)strings we find that the popcorn regime is complementary to the (Gaussian) continuous background. The popcorn signature detection would enhance confidence 
in the Gaussian stochastic background detection and could also help 
determine fundamental string parameters such as the string tension and the reconnection probability. 

In section II we compute the rate of expected burst signals from cosmic string cusps. 
In section III we compute the stochastic background in the popcorn and continuous regimes. 
In section IV we discuss the detection of the two signatures and we compute the expected constraints in the signal parameters space placed by future AdLV, ET, LISA and PTA experiments. Finally, we conclude with a summary of the paper and future research that can be motivated by this work.

\section{Rate of cosmic string bursts and Detection Regimes}
Cusps tend to be formed a few times during each oscillation period \cite{1984NuPhB.242..520T}. The rate of bursts at the observed frequency $f$ from the redshift interval $dz$, from loops of length $l$ is given by \cite{sie06},
\begin{equation}
\frac {dR}{dz}(f,z) = H_0^{-3} \varphi_V(z)(1+z)^{-1} \nu(l,z) \Delta(f,l,z)
\label{eq:dRdz}
\end{equation}
where $H_0$ is the Hubble constant, $\varphi_V(z)$ is the dimensionless co-moving volume element (Appendix A), the factor $(1+z)^{-1}$ corrects 
for the cosmic expansion, and $\nu(z,l)$ is the number of cusps per unit space-time volume from loops with lengths $l$ at redshift $z$. 
Because the GW emission is beamed in the direction of the cusp, only a fraction $\Delta(f,l,z)$ (the beaming fraction) can be observed at frequency $f$. 
When loops formed are small, so that the length is gravitationally radiated away in a Hubble time, $l$ is given by its redshift as:
\begin{equation}
l(z)=\alpha t(z)
\label{eq:lz}
\end{equation}
where $t(z)=H_0^{-1} \varphi_t(z)$ is the Hubble time (Appendix A), $\alpha=\varepsilon \Gamma G\mu$, $G\mu$ is the tension in Planck units ($G$ being the Newton constant), and 
$\Gamma \sim 50$ is a constant related to the power emitted by loops into GWs  \cite{sie06}.

In this case:
\begin{equation}
\nu(z) =\frac{2n_c}{l(z)} n(l(z),z)
\end{equation}
where $n_c$ is the number of cusps per loop oscillation (we will assume $n_c=1$ in average), and $n(z,l)$ is the loop size distribution:
\begin{equation}
n(l,z)=(p \Gamma G\mu)^{-1}t(z)^{-3} \delta(l-\alpha t(z))).
\end{equation}
The beaming fraction is given by:
\begin{equation}
\Delta(f,z) \approx \theta_m^2(f,z) /4 \,\ \mathrm{with} \,\ z<z_m
\end{equation}
where 
\begin{equation}
\theta_m(f,z)= [g_2 (1+z)f l(z)]^{-1/3}
\label{eq:thetam}
\end{equation}
is the maximum angle that the line of sight and the direction of a cusp can
subtend and still be observed at a frequency $f$.
The ignorance constant $g_2$ absorbs factors of O(1), as well as the fraction of the loop 
length $l$ that contributes to the cusp \cite{sie06}. We expect $g_2$ to be of O(1) if loops are smooth.

\begin{figure}
\centering
\includegraphics[angle=0,width=0.49\columnwidth]{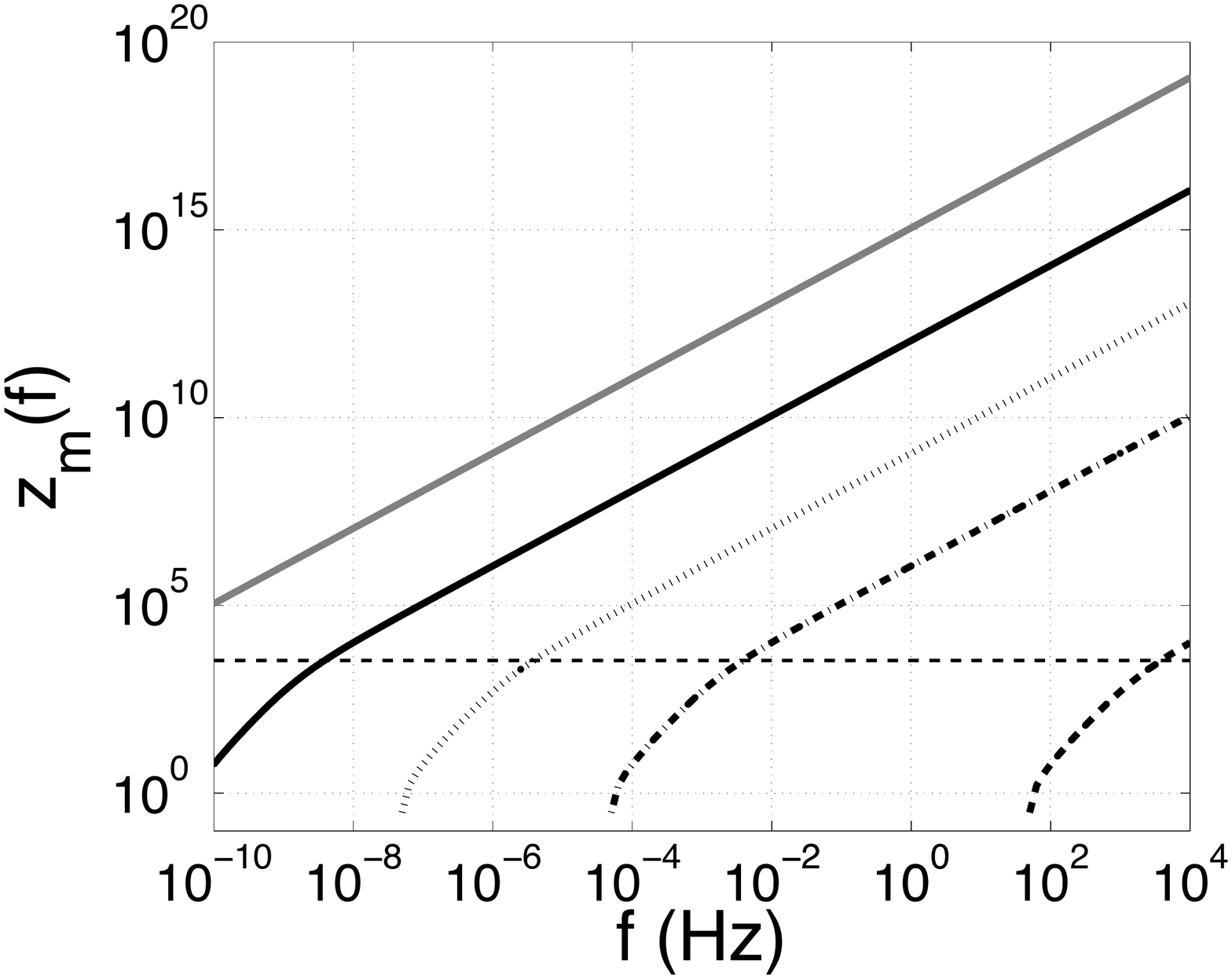}
\includegraphics[angle=0,width=0.49\columnwidth]{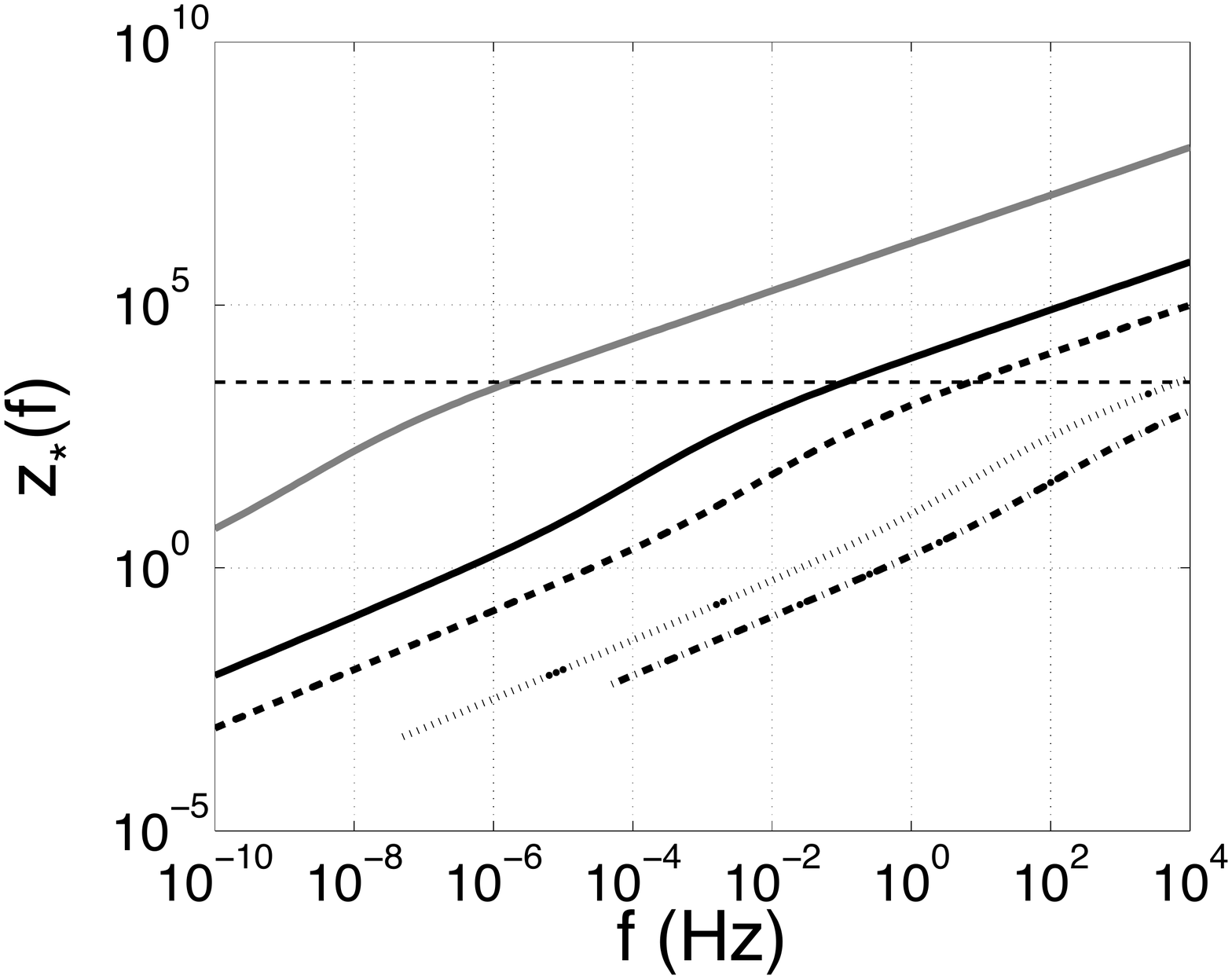}\\
\caption{Left: Maximal contributing redshift as a function the frequency $f$. In the radiation era ($z>z_{eq} \sim 3400$), $z_m(f)=1.25*10^{21} G\mu \varepsilon f$. Right: Transition redshift between popcorn and (Gaussian) continuous regimes ($\Lambda=10$), as a function of the frequency. The grey continuous line corresponds to $p=1$, $\varepsilon=1$ and $G\mu=10^{-6}$, the black continuous line to $p=1$, $\varepsilon=1$ and $G\mu=10^{-9}$,  the black dotted line  to $p=1$, $\varepsilon=1$ and $G\mu=10^{-12}$, the black dashed line  to $p=10^{-3}$, $\varepsilon=1$ and $G\mu=10^{-9}$, and the black dot-dashed line  to $p=1$, $\varepsilon=10^{-6}$ and $G\mu=10^{-9}$. The vertical dashed line indicates the transition redshift between matter and radiation eras.
\label{fig:zm}}
\end{figure}

\begin{figure}
\includegraphics[angle=0,width=0.49\columnwidth]{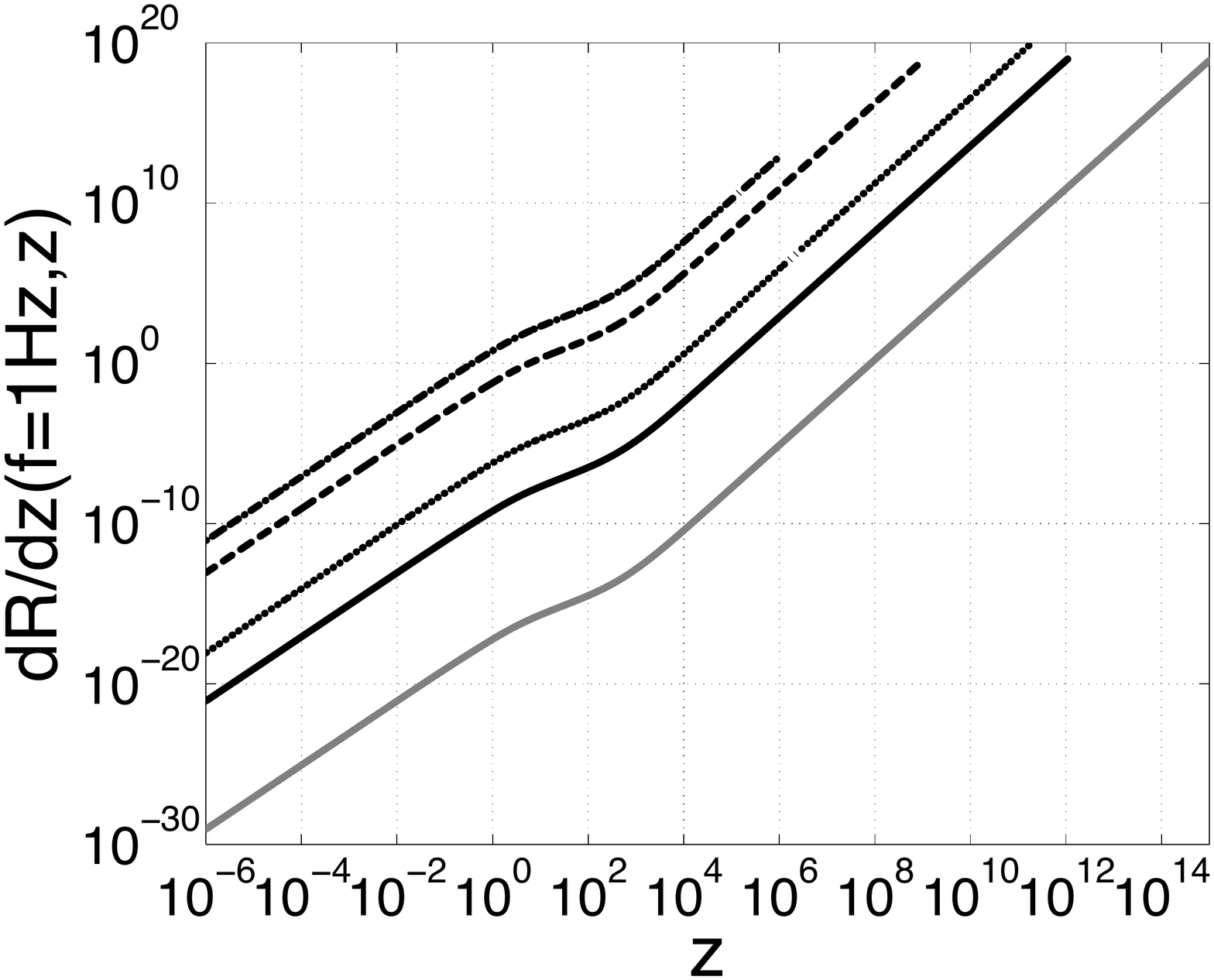}
\includegraphics[angle=0,width=0.49\columnwidth]{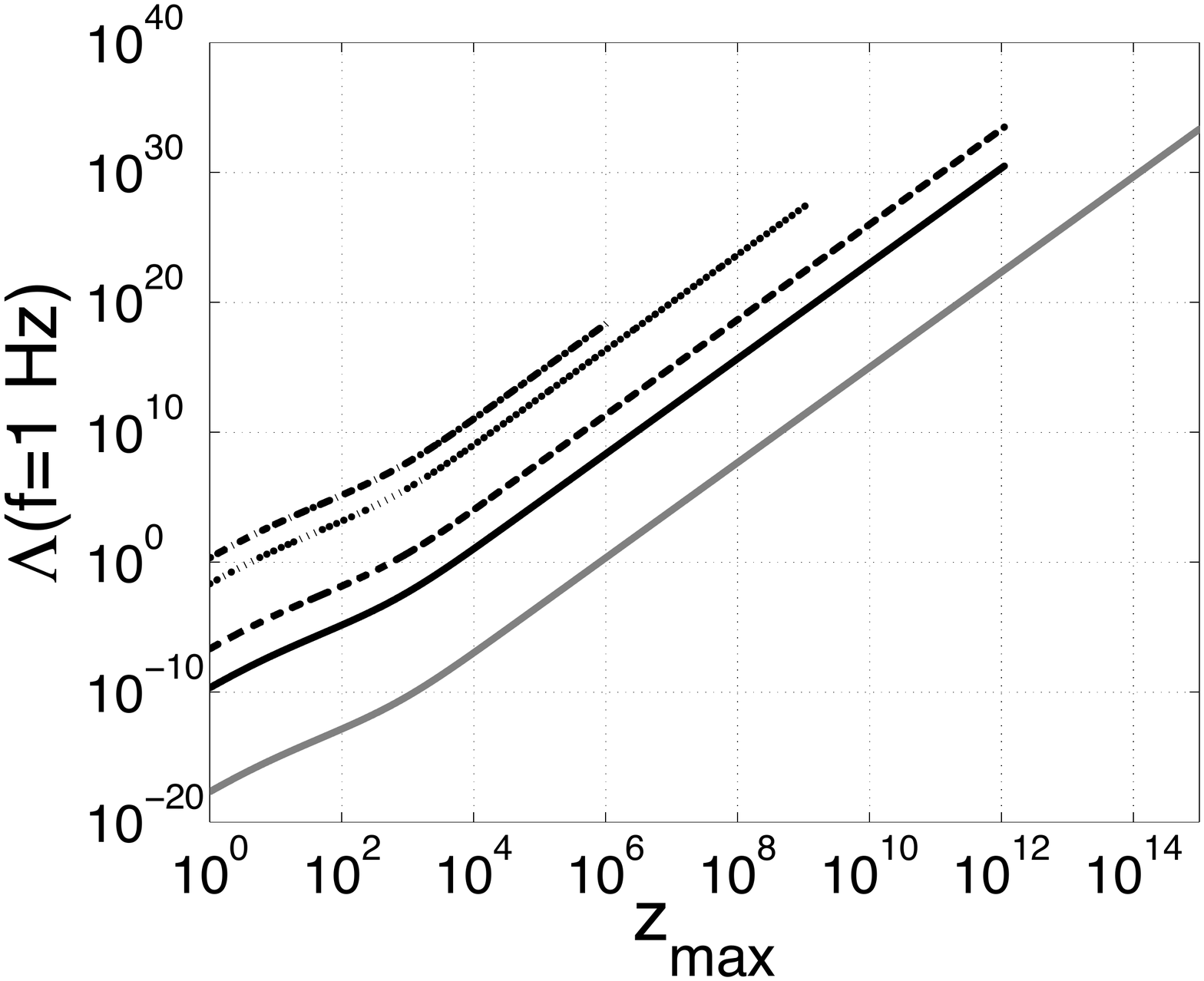}
\caption{Left: Rate of cosmic strings per interval of redshift at the frequency $f=1$ Hz. Right: Average number of  bursts overlapping at the frequency $f=1$ Hz, as a function of the maximal redshift $z_{\max}$.
For other frequencies, the rate can be deduced from these plots 
by multiplying $dR/dz$ by $f^{2/3}$. The grey continuous line corresponds to $p=1$, $\varepsilon=1$ and $G\mu=10^{-6}$, the black continuous line to $p=1$, $\varepsilon=1$ and $G\mu=10^{-9}$, the black dotted line to $p=1$, $\varepsilon=1$ and $G\mu=10^{-12}$, the black dashed line to $p=10^{-3}$, $\varepsilon=1$ and $G\mu=10^{-9}$, and the black dot-dashed line to $p=1$, $\varepsilon=10^{-6}$ and $G\mu=10^{-9}$.
\label{fig:dRdz}}
\end{figure}

The cutoff redshift is solution of the equation $\theta_m (f,z_m)=1$ and is shown in Fig.~\ref{fig:zm}. 
Interestingly, the dependence on the string parameters $G\mu$, $\varepsilon$, $p$ and on the frequency $f$ is going to enter through a pre-factor, and we can then  re-write the expression of the rate as:
\begin{equation}
\frac {dR}{dz}(f,z)  = B(f) F(z) \,\ \mathrm{with} \,\ z<z_m(f)
\end{equation}
with
\begin{equation}
F(z) =\varphi^{-14/3}_t(z) \varphi_V(z) (1+z)^{-5/3}
\end{equation}
and 
\begin{eqnarray}
B(f) = \frac{g_2^{-2/3}n_c H_0^{5/3}f^{-2/3}}{2 \Gamma^{8/3}(G\mu)^{8/3} \varepsilon^{5/3}p}.
\end{eqnarray}
The rate as a function of the redshift $z$ for different sets of parameters and for a frequency $f=1$ Hz is plotted in Figure~\ref{fig:dRdz}. 
The rate increases with $z$ until it reaches the cutoff $z_m$, with a slope which is larger in the radiation era after $z_{eq} \sim 3400$ (Appendix A). 
As expected, the rate is larger for smaller values of the reconnection probability $p$, the tension $G \mu$ or the typical length $\varepsilon$.

The GW signal from the population of cosmic strings falls into three different statistical regimes, characterized by the value of the quantity (see Figure~\ref{fig:dRdz}):  
\begin{equation}
\Lambda(f,z)= \tau \int_0^z \frac{dR}{dz}(f,z) dz
\end{equation}
where $\tau$ is the duration of the signal (typically $1/f$ at the frequency $f$ \cite{dam05,buo05} ), and the integral is the inverse of the time interval between successive events arriving from redshift $<z$. This quantity is often called the {\it duty cycle} \cite{buo05} and is simply the average number of sources overlapping in a typical frequency band $\sim f$, around the frequency $f$. It can be compared to the overlap function of \cite{ros11}, which is the number of sources present, in average, in a frequency bin $\Delta f$ around the frequency $f$. 
Notice that taking $\Delta f \sim 1/T$, where $T$ is the observation time, determines whether sources create a {\it confusion background} in the framework of single source detection \cite{ros11}, which is not the purpose of this work, where we are interested in the detection of the background itself.
 

Based on the value of $\Lambda$ we distinguish the following three regimes:
\begin{enumerate}
\item {\bf shot noise}: at low redshift,  the number of sources is small enough for the time interval between events to be long compared to the duration of a single event. 
Sources are separated by long stretches of silence and the closest ones may be detected individually.
\item {\bf popcorn noise}: when the redshift and the number of sources increases, the time interval between events becomes comparable to the duration of a single 
event. The number of sources present at the frequency $f$ is a Poisson process, sometimes there is no GW signal, sometimes sources overlap, and the sum of the amplitudes at a given time is still unpredictable. These signals, which sound like crackling popcorn, are known as {\it popcorn noise}. 
\item { \bf continuous}: the number of sources is large enough for the time interval between events to be small compared to the duration of a single event. Sources overlap at the frequency $f$ to create a continuous background (there is always a GW signal present) that is Gaussian in nature (due to the central limit theorem, if the number of sources is large, the sum of their amplitudes has a Gaussian distribution) and can be confounded with the detector noise. 
\end{enumerate}

\cite{dam00,dam01,dam05} first discussed the presence of a popcorn-like noise on top of a continuous stochastic background for cosmic strings . 
They adopted the value of $\Lambda=1$ as the limit between the continuous and popcorn 
regime. In this paper we follow \cite{cow06} and consider a 
more conservative value of $\Lambda=10$, in order to ensure also Gaussianity. It is worth pointing out that the results are not too sensitive to this choice.

\section{The stochastic background}

The spectrum of the gravitational stochastic background is usually
characterized by the dimensionless parameter \cite{all99}:
\begin{equation}
\Omega_{gw}(f)=\frac{1}{\rho_c}\frac{d\rho_{gw}}{d\ln f}
\end{equation}
and for the case of cosmic strings is given by\cite{sie07}: 
\begin{equation}
\Omega_{gw}(f) =  \frac{4 \pi^2}{3H_0^2} f^3  \int_{z_0}^{z_m(f)} \tilde{h}^2(f,z)  \frac {dR}{dz}(f,z) dz
\end{equation}
where we have added up individual contributions at all redshifts, excluding the very nearby ($z_0 \approx 10^{-6}$ or 5 kpc), and where the gravitational strain produced by a cosmic string cusp is~\cite{dam00,dam01,dam05,sie06,sie07},
\begin{equation}
\tilde{h}(f) = A f^{-4/3} \Theta(f_h-f) \Theta(f-f_l).
\label{eq:hf}
\end{equation}
The low frequency cutoff of the gravitational wave signal, $f_l$, is determined by the size of the cusp -- a scale which is typically 
cosmological. As a result the low frequency cutoff of detectable radiation is determined by the low frequency behavior of the 
instrument (for terrestrial detectors, for instance, by seismic noise). The high frequency cutoff depends on the angle between the 
line of sight and the direction of the cusp $\theta$, as
\begin{equation}
f_h = [l(z)(1+z)\theta^3]^{-1}.
\label{eq:fh}
\end{equation}

The amplitude $A$ of a cusp from a loop of length $l$ at a redshift $z$ is given by~\cite{sie06},
\begin{eqnarray}
A & =& g_1 \frac{G\mu l^{2/3} H_0} {(1+z)^{1/3}  \varphi_r (z)}.\\
& = &\frac{g_1 \Gamma^{2/3} H_0^{1/3} (G\mu)^{5/3} \varepsilon^{2/3} \varphi_t(z)^{2/3}} {(1+z)^{1/3}  \varphi_r (z)}.
\label{eq:ampl}
\end{eqnarray}
Here $g_1$ is an ignorance constant that absorbs the uncertainty on exactly how much of the length $l$ is involved in the production of 
the cusp and $\varphi_r (z)$ is the dimensionless proper distance (Appendix A).

We define the stochastic background as the sum of the continuous and popcorn contributions:
\begin{equation}
\Omega_{gw}(f)=\Omega^{pop}_{gw}(f) + \Omega^{cont}_{gw}(f)
\end{equation}
where
\begin{equation}
\Omega^{pop}_{gw}(f) = \frac{4 \pi^2}{3H_0^2} f^3 \int_{z_0}^{z*(f)}  \tilde{h}^2(f,z)  \frac {dR}{dz}(f,z) dz
\label{eq:omegapop}
\end{equation}
and
\begin{equation}
\Omega^{cont}_{gw}(f) = \frac{4 \pi^2}{3H_0^2} f^3 \int_{z*(f)}^{z_m(f)} \tilde{h}^2(f,z)  \frac {dR}{dz}(f,z) dz
\label{eq:omegacont}
\end{equation}
where $z_*(f)$ is the redshift at which $\Lambda(z_*(f))=10$ (see Fig.~\ref{fig:zm})

The shape of the continuous background is determined by a very sharp rise at $f_0 \approx 4.7 \times 10^{-20} \varepsilon^{-1} (G\mu)^{-1}$ where sources at low redshift start to become observable ($\theta_m(f_0,z_0)=1$), followed by a decrease ($\sim f^{-1/3}$), and a flat region where most of the sources belong to the radiation era and where the increase of the rate with frequency cancels the decrease of their contribution to $\Omega_{gw}$. The popcorn background has a similar behavior, except for the flat part of the spectrum. It raises more smoothly at lowest frequencies because the upper limit of the integral of Eq.~\ref{eq:omegapop} (the transition redshift $z_*$) increases slower with frequency than $z_m$. Near $f_0$ there is a very small region where the popcorn background dominates as $z_m$ is still in the popcorn regime or very close to the transition redshift so that there is an 
insignificant continuous contribution. Sometimes this region is not be visible in the plots due to our numerical precision. 
\begin{figure}
\includegraphics[angle=0,width=\columnwidth]{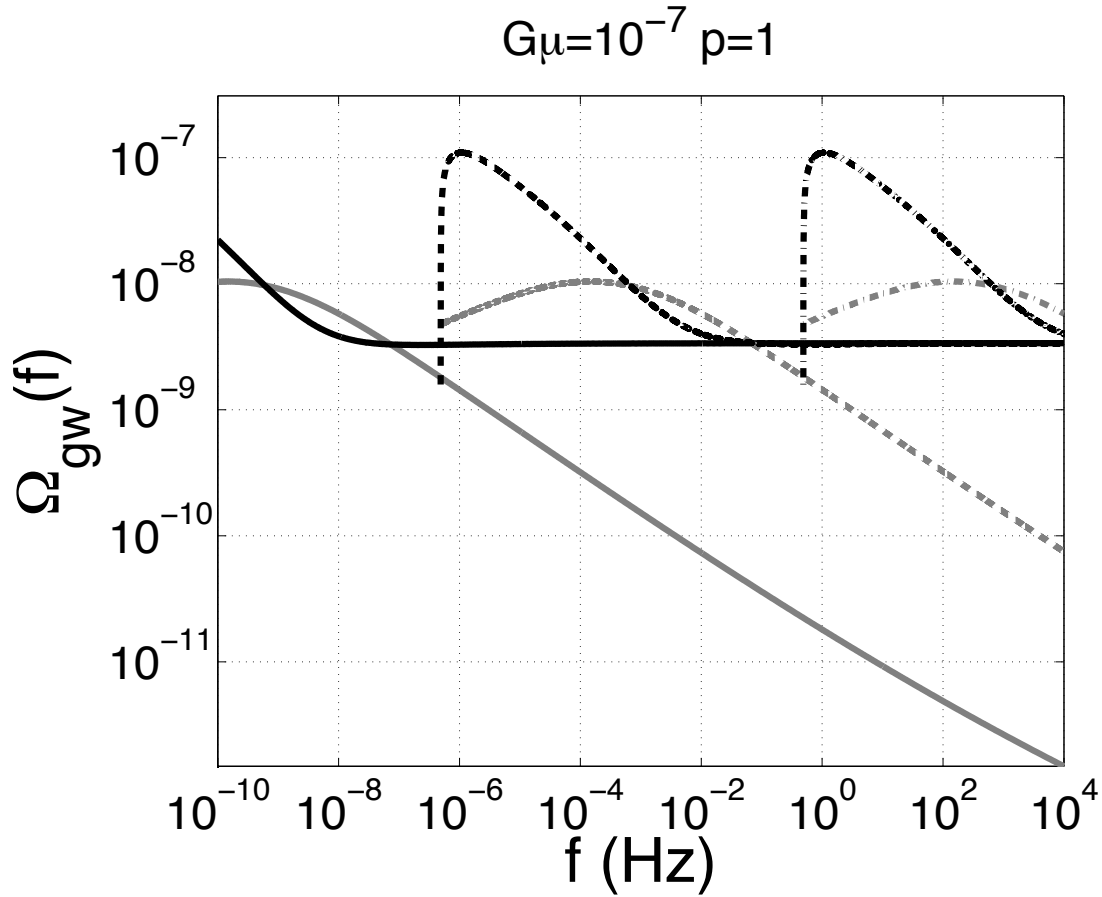}
\caption{Popcorn (grey) and continuous (black) contributions to $\Omega_{gw}(f)$ for $p=1$, $G\mu=10^{-7}$ and $\varepsilon=1$ (continuous line), $10^{-6}$ (dashed line) and  
$10^{-12}$ (dot-dashed line) . Increasing $\varepsilon$ shifts the spectra toward lower frequencies but doesn't affect the amplitude. \label{fig:omega_eps}}
\end{figure}
\begin{figure}
\centering
\includegraphics[angle=0,width=\columnwidth]{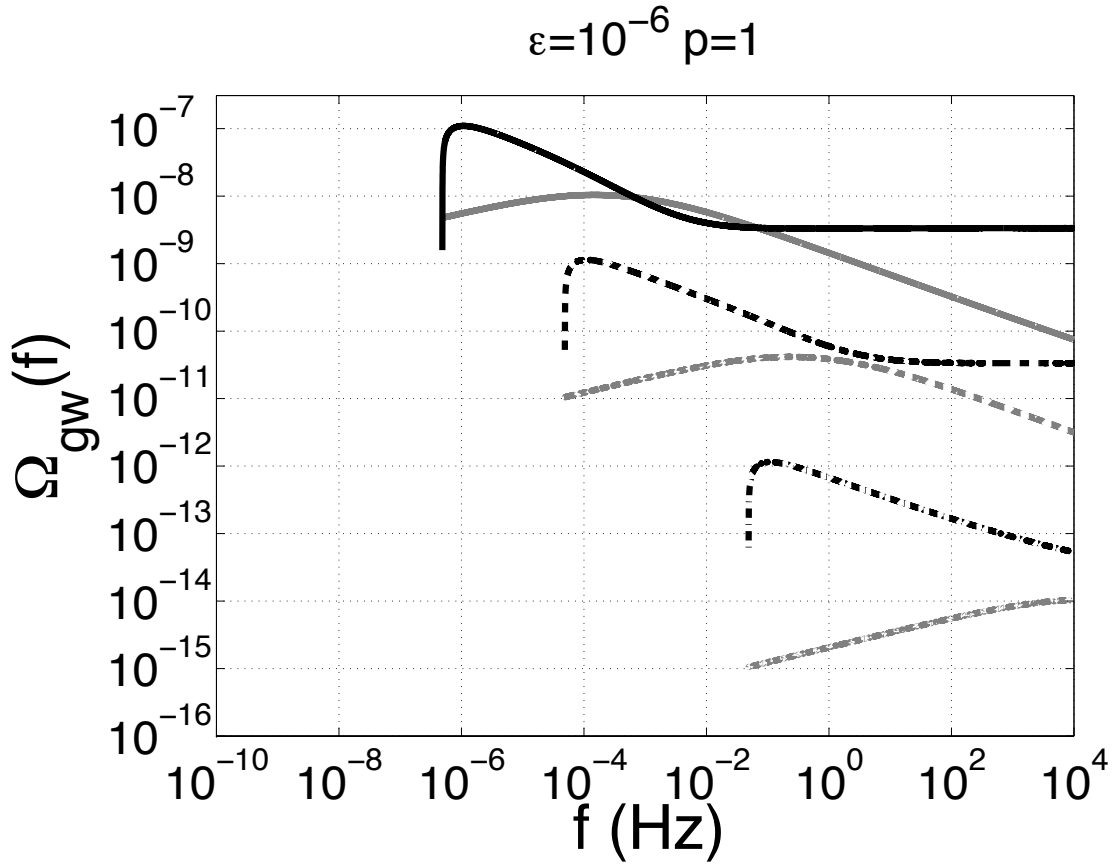}
\caption{Popcorn (grey) and continuous (black) contributions to $\Omega_{gw}(f)$ for $p=1$, $\varepsilon=10^{-6}$ and $G\mu=10^{-7}$ (continuous line), $10^{-9}$ (dashed line) 
and $10^{-12}$ (dot-dashed line). Increasing $G\mu$ shifts the spectra toward lower frequencies, increases the amplitude and the relative importance of the continuous contribution 
compared to the popcorn contribution.
\label{fig:omega_Gmu}}
\end{figure}
\begin{figure}
\centering
\includegraphics[angle=0,width=\columnwidth]{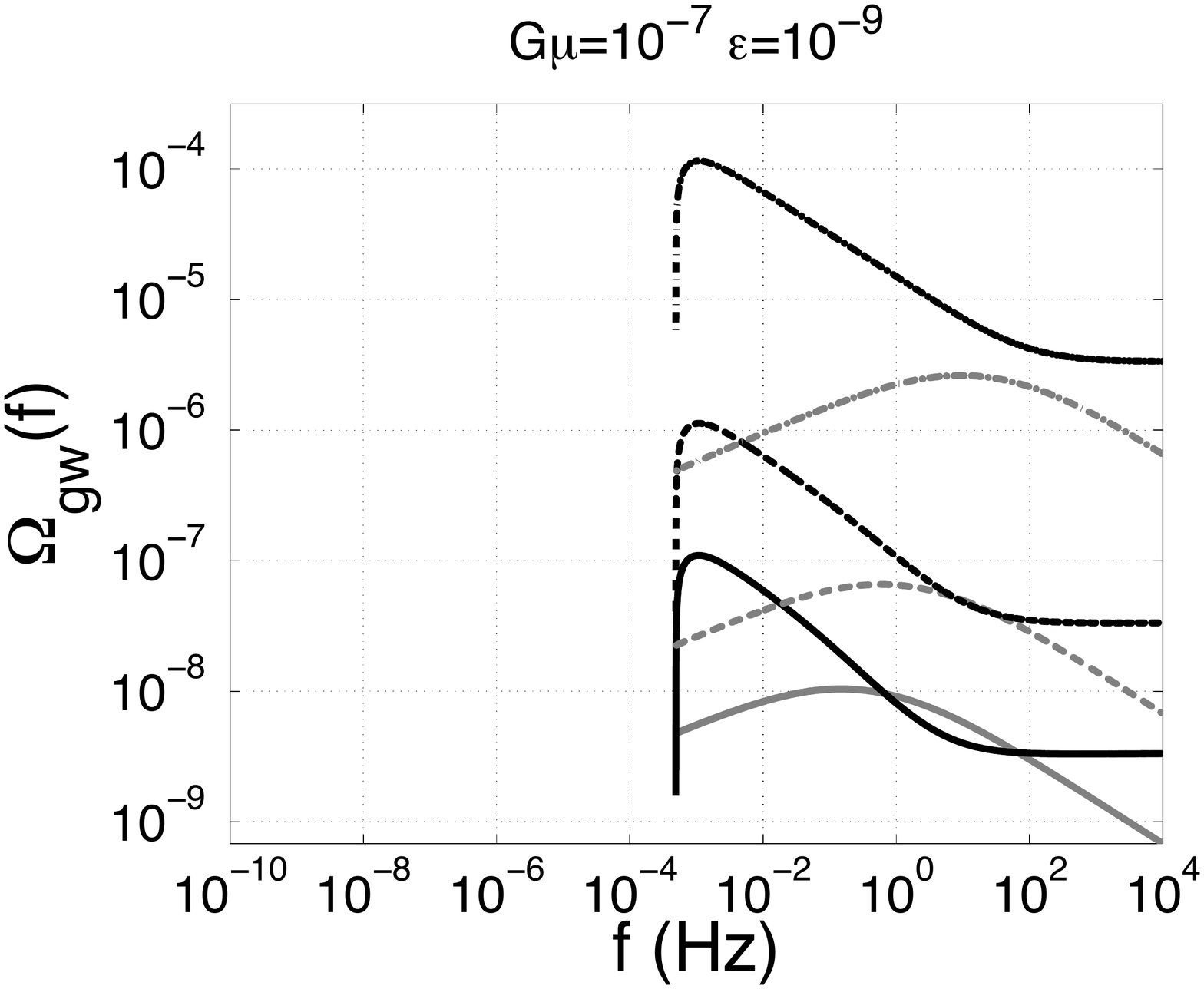}
\caption{Popcorn (grey) and continuous (black) contributions to $\Omega_{gw}(f)$ for $G\mu=10^{-7}$, $\varepsilon=10^{-9}$ and $p=1$ (continuous line),  $0.1$ (dot-dashed line) 
and $10^{-3}$ (dot-dashed line). Increasing $p$ shifts the popcorn spectrum toward lower frequencies, decreases the amplitude of both the continuous and popcorn spectra and the 
relative importance of the continuous contribution compared to the popcorn contribution.
\label{fig:omega_p}}
\end{figure}
Figures~\ref{fig:omega_eps},~\ref{fig:omega_Gmu} and~\ref{fig:omega_p} compare the two contributions for different sets of parameters. Depending on the 
parameters, the popcorn regime may overwhelm the continuous background in the frequency range of AdLV, ET, LISA and PTA.
The parameters $G \mu$, $\varepsilon$ and $p$ affect the amplitude of the power spectra through an overall scaling factor $F_s \propto \varepsilon^{-1/3} (G\mu)^{2/3}p^{-1}$  but also through the bounds $z_m \propto \varepsilon G\mu$ and $z_*$ of the integrals over redshift. In the case of $\varepsilon$ the two effects compensate so that finally, the amplitude doesn't depend on this parameter. Another interesting feature is that the parameter $p$ doesn't affect the two backgrounds the same way. In fact, unlike $z_m(f)$, the popcorn maximal redshift depends on the parameter $p$. Decreasing the reconnection probability $p$ increases the overall factor $F_s$ but also reduces $z_*$, canceling part of the gain. 
 
\begin{figure}
\centering
\includegraphics[angle=0,width=0.45\columnwidth]{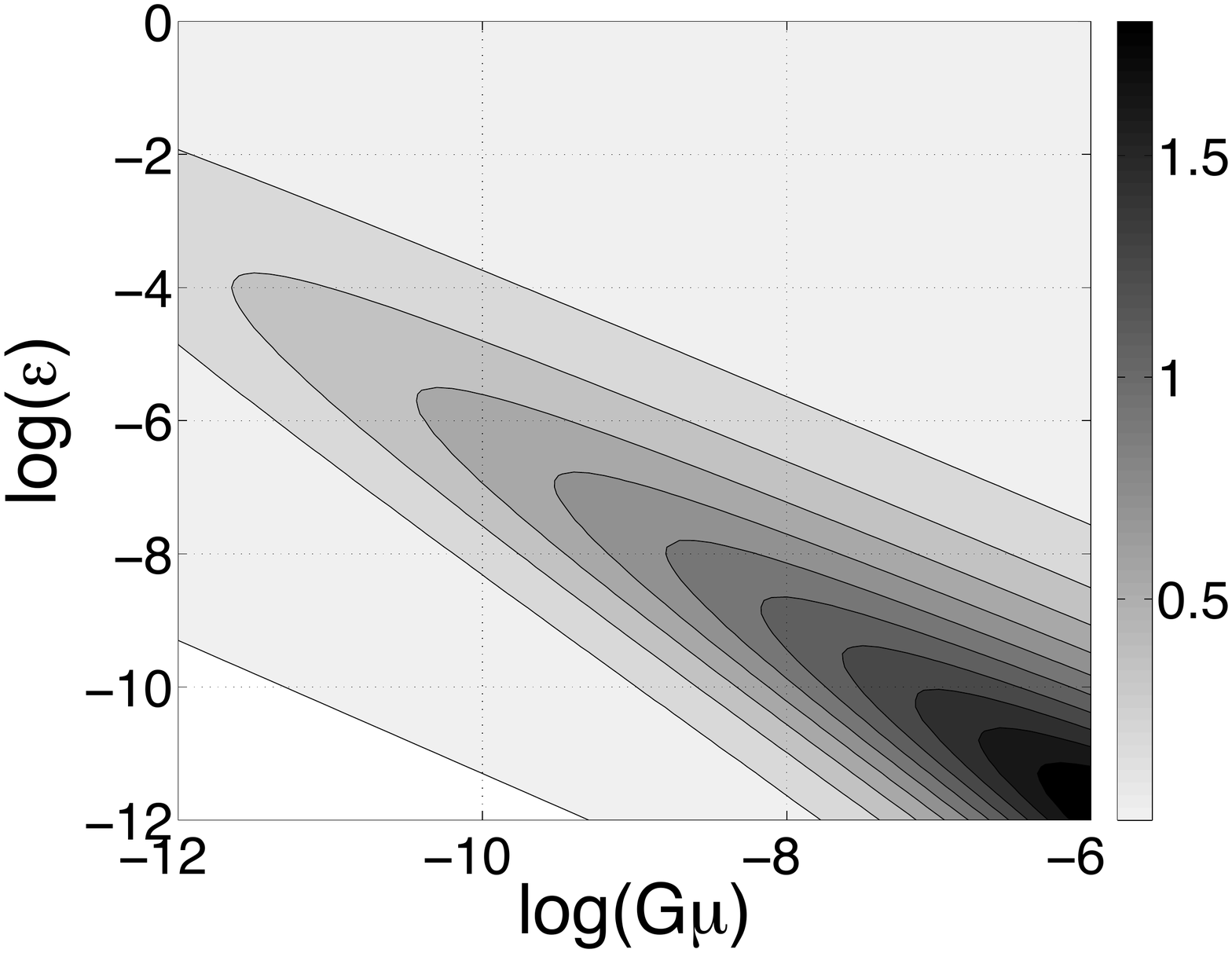}
\includegraphics[angle=0,width=0.45\columnwidth]{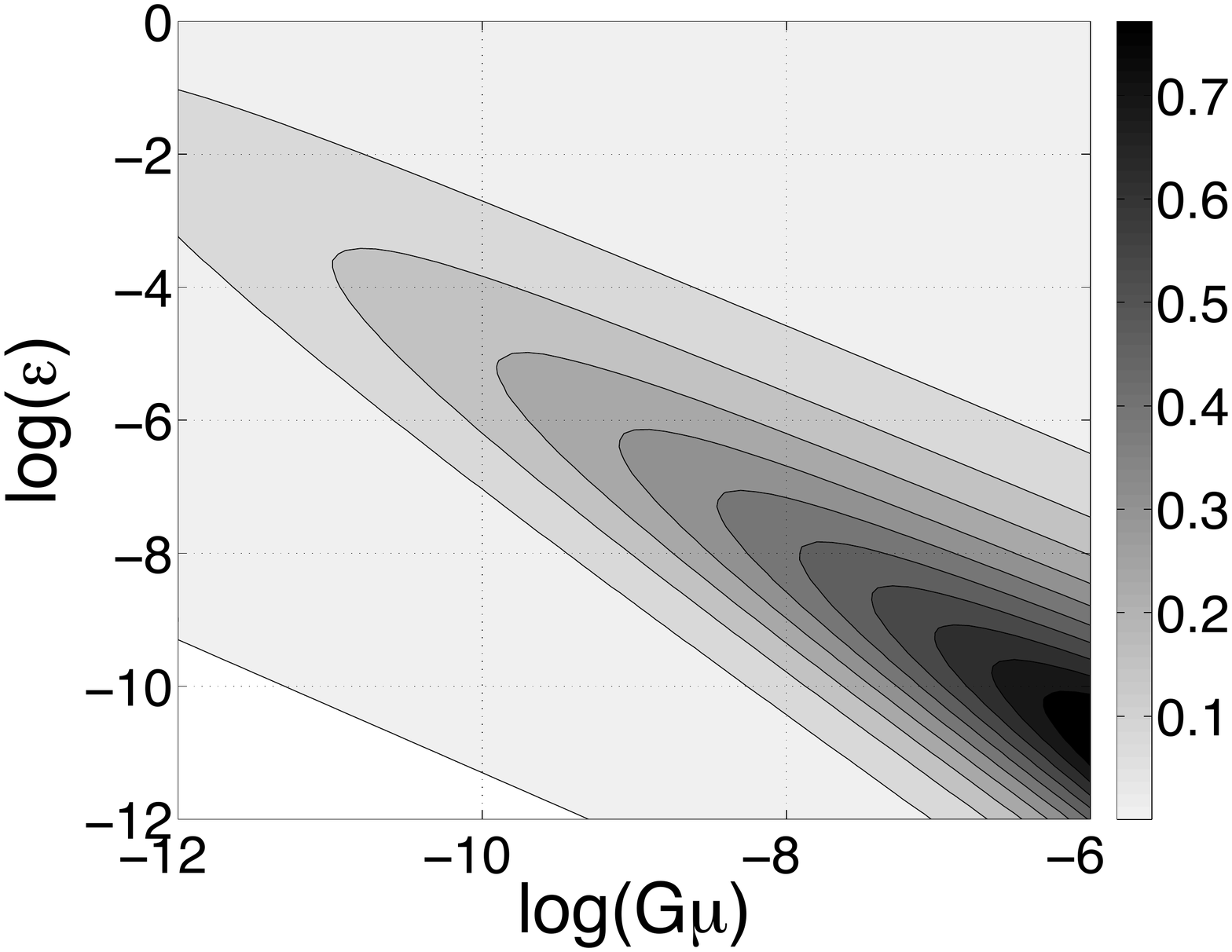}\\
\includegraphics[angle=0,width=0.45\columnwidth]{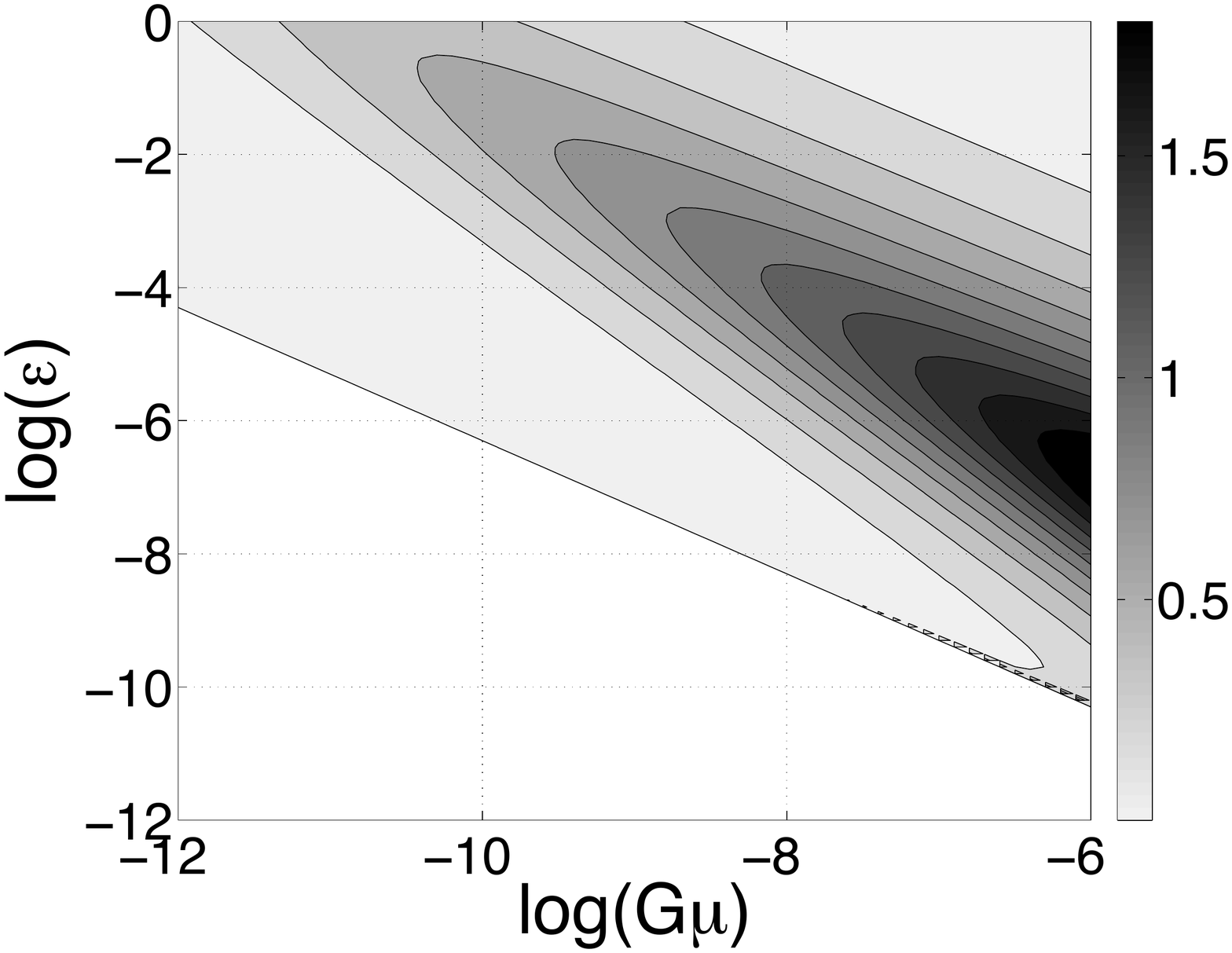}
\includegraphics[angle=0,width=0.45\columnwidth]{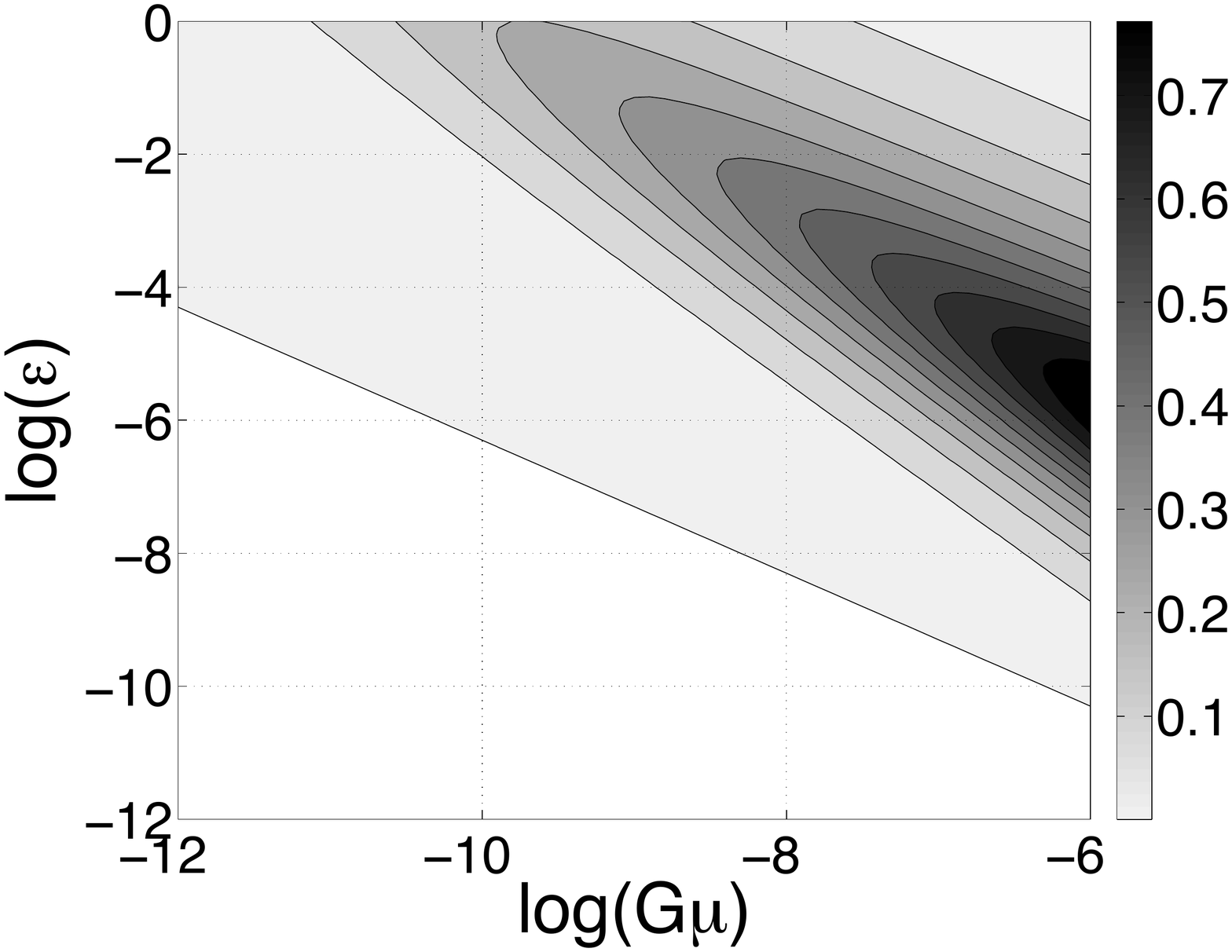}\\
\includegraphics[angle=0,width=0.45\columnwidth]{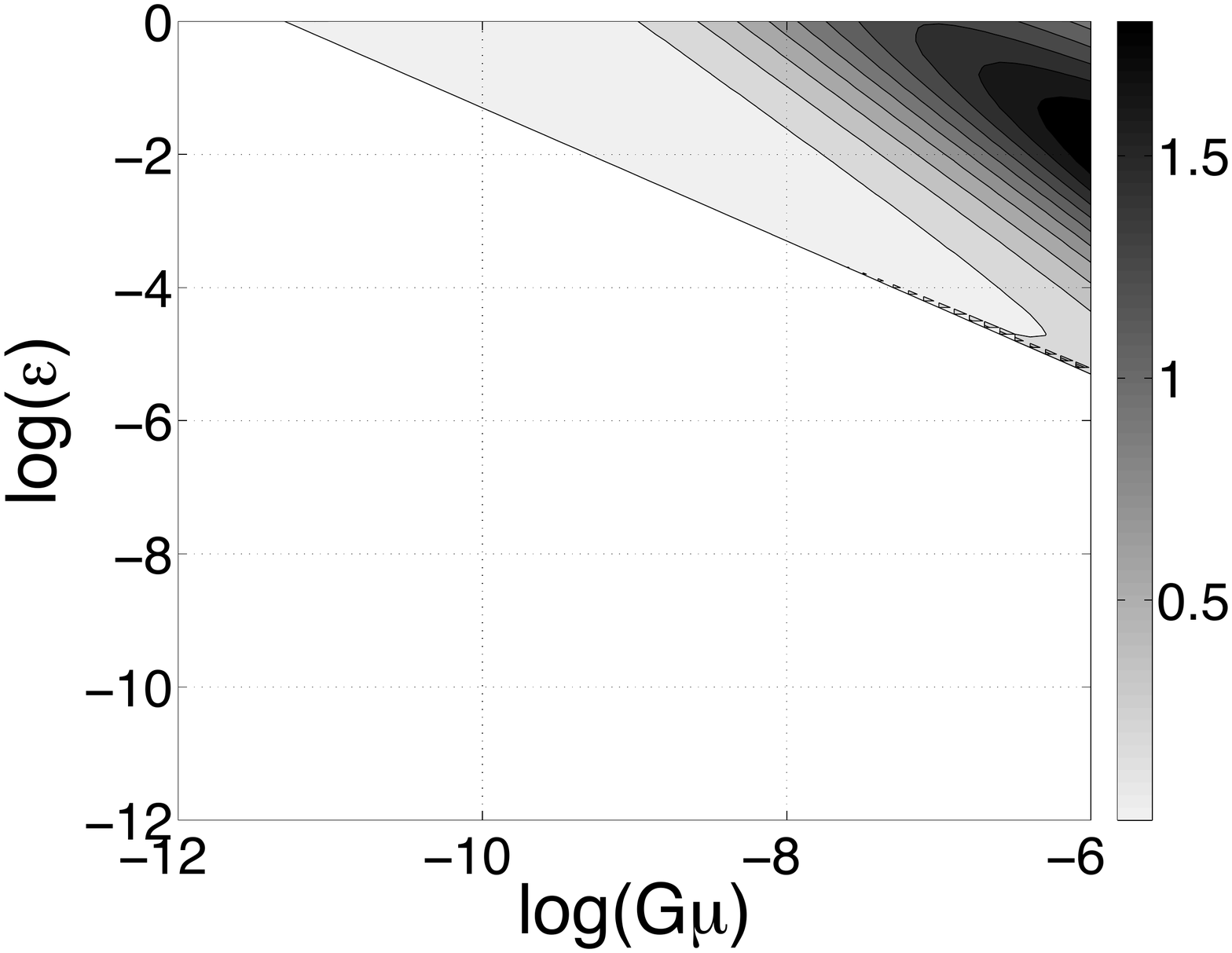}
\includegraphics[angle=0,width=0.45\columnwidth]{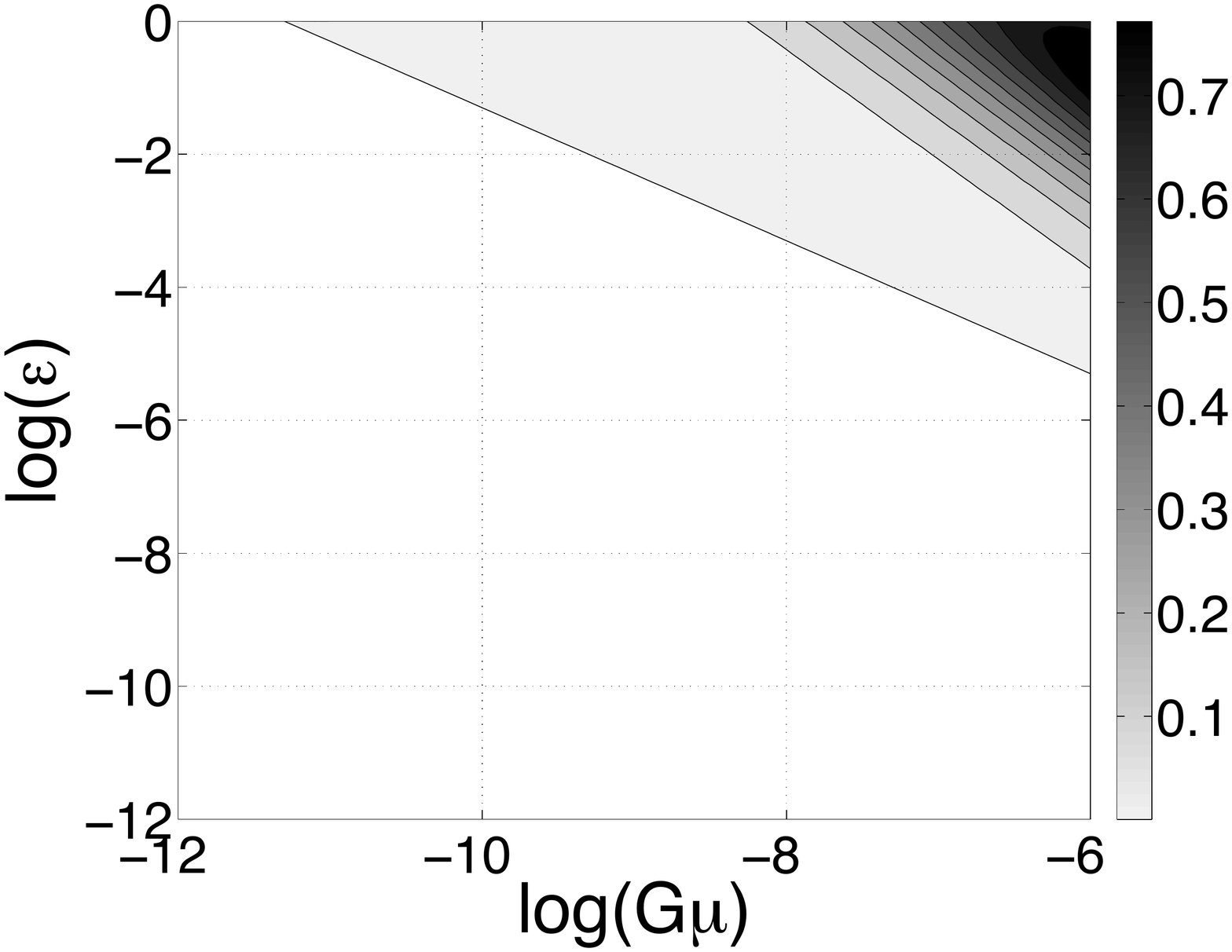}\\
\caption{ratio $R=\Omega_{gw}^{pop} / \Omega_{gw}^{cont}$ for $p=1$ (left) and for $p=10^{-3}$ (right), and from top to bottom at $f=100$, $10^{-3}$ and $10^{-8}$ Hz. For $p=1$, in the denser regions from dark black to grey, the popcorn background contribution dominates and is expected to significantly contribute to the signal-to-noise ratio (see section IV). For $p=10^{-3}$, on the other hand the background is always dominated by the  continuous contribution. Notice that the greyscale color bar has different scales in the two cases.
The small sharp area at the right bottom of the two last plots of the first column is not an artefact, but corresponds to the lower frequency part of the spectrum where the number of sources is larger in the popcorn regime than in the continuous regime (see discussion in section III).  
In the white region in the bottom left corner, both the popcorn and the continuous backgrounds are null ($\Omega_{gw}=0)$.}
\label{fig:ratio}
\end{figure}

Figure~\ref{fig:ratio}  shows the corresponding regions in the plane $\varepsilon$-$G\mu$ with constant ratio $R=\Omega_{gw}^{pop}/\Omega_{gw}^{cont}$, for frequencies 
$f=100, 10^{-3}$ and $10^{-8}$ Hz, typical for terrestrial detectors (AdLV, ET), LISA and PTA, and for two different values of the parameter 
$p$ ($p=1$ and $p=10^{-3}$). For $p=1$, in the denser regions from dark black to heavy grey, the popcorn background contribution dominates and is expected to significantly contribute to the 
signal-to-noise ratio (see section IV). For $p=10^{-3}$,  the background is always dominated by the  continuous contribution.
In the complementary light grey regions the continuous contribution dominates. Decreasing the frequency, the popcorn dominated area is shifted toward larger values of 
$\varepsilon$. For the LISA typical frequency $f=10^{-3}$, the popcorn dominated area is slightly reduced, while for the PTA frequency $f=10^{-8}$, most of this region 
disappears outside the range of values for $\varepsilon$, so that the continuous background dominates in almost all of the parameter space, except for the largest 
values of both $\varepsilon$ and $G \mu$. The effect of decreasing $p$ is similar to decreasing $f$ and the region where the popcorn contribution dominates is even 
smaller at LISA and  PTA frequencies.

\section{Parameter space Constraints}
In this section we concentrate on the detection of the stochastic background with planned/proposed future detectors and the bounds they can place in the parameter space. 
For a given detector we assume that the gravitational wave emission from a single cosmic string cusp is present for approximately $\tau=1/f_L$, where $f_L$ is the 
lowest observable detector frequency. We then compute the popcorn and  continuous contributions to the stochastic background using Eq.~\ref{eq:omegapop} 
and \ref{eq:omegacont}, requiring that $\Lambda(f)$ is smaller or larger than 10.

For  terrestrial interferometers (AdLV, ET) we use the frequency domain method of cross-correlation between pairs of detectors \cite{all99}. This technique has been shown to be optimal for continuous stochastic backgrounds and to perform nearly optimally for the popcorn contribution 
down to very small values of $\Lambda$ \cite {dra03} (much smaller than our threshold value of $\Lambda = 10$). The signal-to-noise ratio ($SNR$), for an integration time $T$, obtained by cross-correlating two interferometers, is given by \cite{all99}:
\begin{equation}
SNR =\frac{3 H_0^2 F^2_{\mathrm{ifo}}}{4 \pi^2} \sqrt{2 T} \left[ \int_0^\infty
df \frac{\gamma^2(f)\Omega_{\rm gw}^2(f)}{f^6 P_1(f)P_2(f)} \right]^{1/2}
\label{eq:snrCC}
\end{equation}
where $\gamma$ is the normalized overlap reduction function characterizing the loss in sensitivity due to the separation and the relative orientation of the detectors, 
$F^2_{\mathrm{ifo}}=<F^2_{+}+F^2_{\times}>=2\sin^2(\alpha)/5$ is the sum of the detector power pattern functions, averaged over sky position and polarization 
(giving a measure of the angular efficiency), $\alpha$ is the opening angle between the interferometer's arms ($\pi/2$ for AdLV, $\pi/3$ for ET), and $P_1$ and $P_2$ 
are the strain noise power spectral densities of the two detectors. 
 
For LISA and PTA we simply compare the GW signal to the expected sensitivity (using the LISA strain noise power spectral density of the standard Michelson configuration~\cite{LISA} and the Parkes PTA's projected sensitivity ~\cite{Jenet06}).
It is worthwhile to notice that for LISA it may be possible to combine the symmetrized Sagnac with the Michelson configuration and nearly achieve the sensitivity of cross correlating two LISA detectors~\cite{hogan01,vec02}. However giving the uncertainties on the planned configuration, we preferred not to consider it in this paper.

\begin{figure}
\centering
\includegraphics[angle=270,width=1\columnwidth]{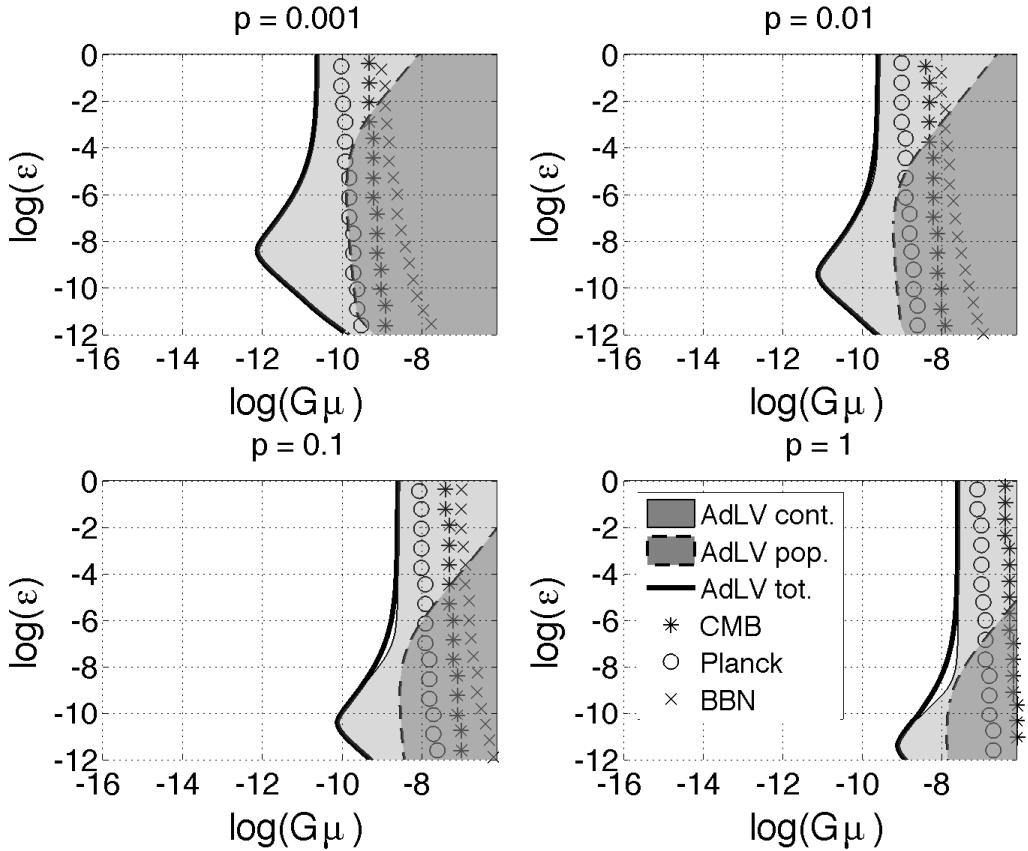}
\caption{AdLV sensitivity (at $2 \sigma$ level) in the plane $((G\mu))-\varepsilon$ for string parameters $p =$ 0.001, 0.01, 0.1, 1. The continuous and popcorn contributions, in light and darker grey colors respectively, 
represent the part of the parameter space that is expected to be probed in a search for a stochastic background using a pair of coincident and co-located AdLV detectors and integrating over a year.
The solid dark line indicates the sensitivity region (to the right of the curve) arising from the sum of the popcorn and continuous contribution. For comparison we 
also show the Big Bang Nucleosynthesis (BBN) and CMB observation limits on $\Omega$, including the future Planck experiment \cite{sie07,S5} on the GW stochastic background formed before BBN ($z \sim 5.5 \times 10^{9}$) and before the CMB photons decoupled ($z \sim 1100$) respectively.
\label{fig:sensitivity1}}
\end{figure}

\begin{figure}
\centering
\includegraphics[angle=270,width=1\columnwidth]{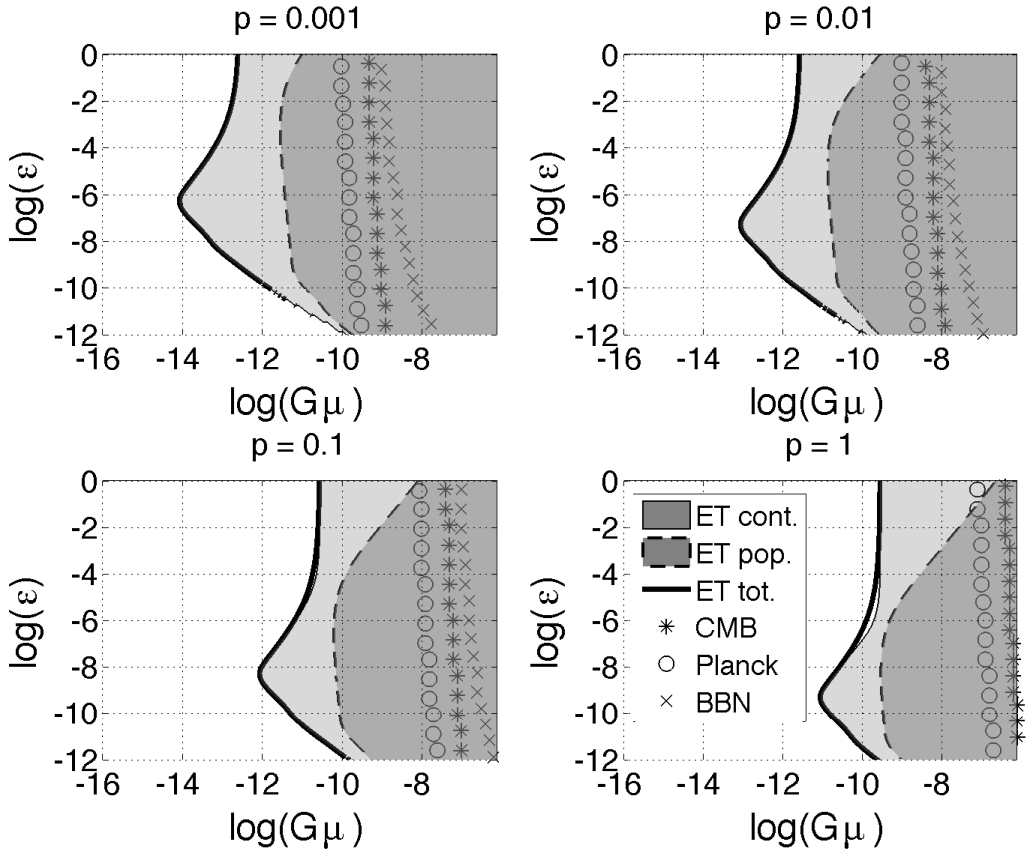}
\caption{ET-B (ET in broadband sensitivity configuration) sensitivity (at $2 \sigma$ level) in the plane $((G\mu))-\varepsilon$ for string parameters $p =$ 0.001, 0.01, 0.1, 1.
\label{fig:sensitivity2}}
\end{figure}

\begin{figure}
\centering
\includegraphics[angle=270,width=1\columnwidth]{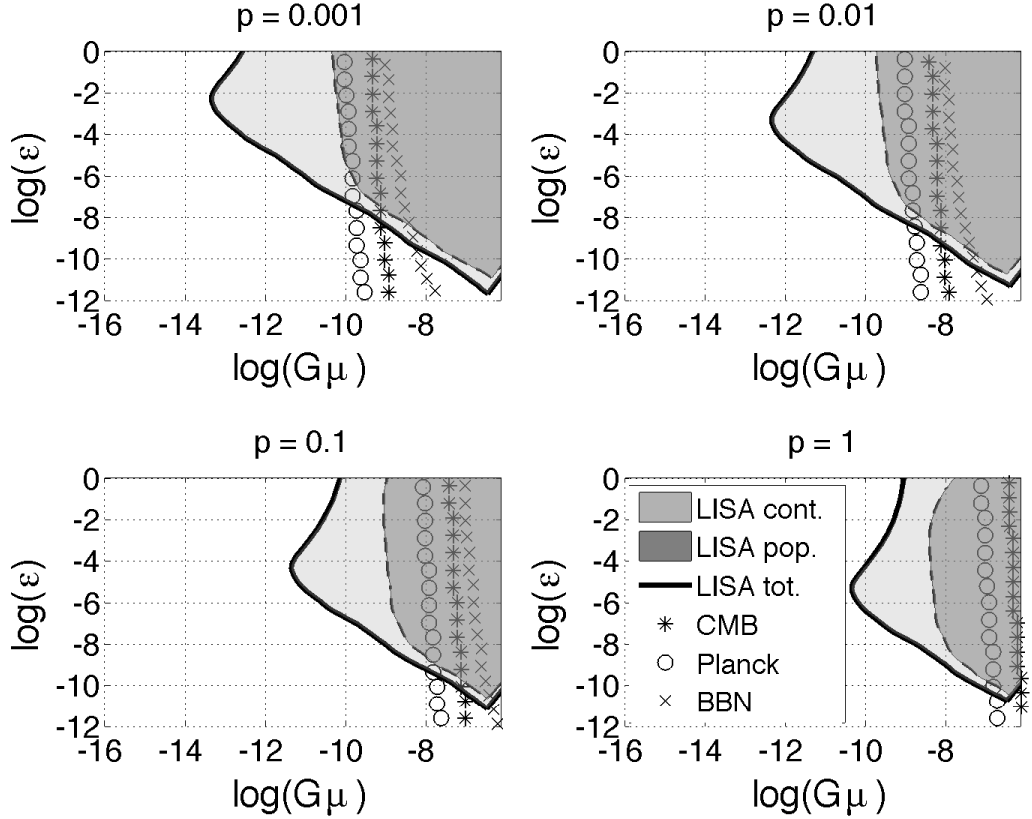}
\caption{LISA sensitivity (at $2 \sigma$ level) in the plane $((G\mu))-\varepsilon$ for string parameters $p =$ 0.001, 0.01, 0.1, 1.
\label{fig:sensitivity3}}
\end{figure}

\begin{figure}
\centering
\includegraphics[angle=270,width=1\columnwidth]{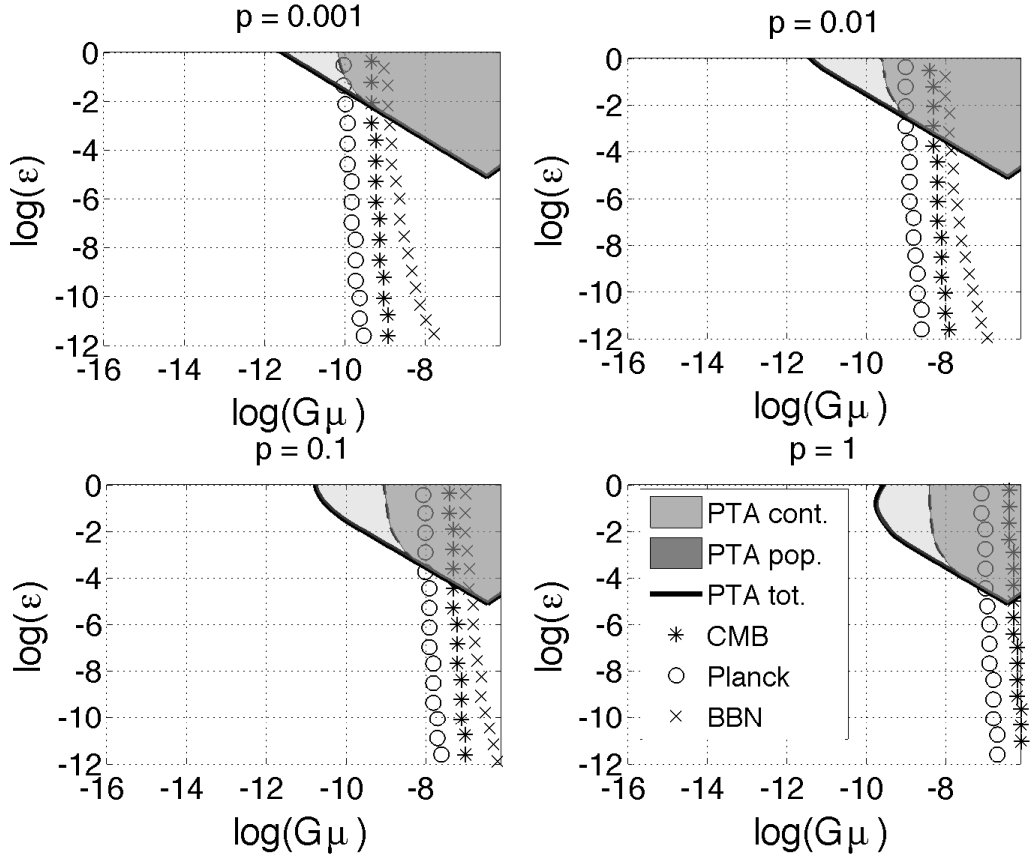}
\caption{PTA sensitivity in the plane $((G\mu))-\varepsilon$ for string parameters $p =$ 0.001, 0.01, 0.1, 1. 
PTA's sensitivity in $\Omega_{gw}$ corresponds to a $0.1\%$ false alarm rate and a $95\%$ detection rate upper bound (using~\cite{Jenet06} table 3 for $\alpha=-1$), equivalent to $2 \sigma$ level.
\label{fig:sensitivity4}}
\end{figure}

Figures~\ref{fig:sensitivity1},~\ref{fig:sensitivity2},~\ref{fig:sensitivity3} and~\ref{fig:sensitivity4} show regions in the $\varepsilon$-$G\mu$ plane where both the popcorn 
and the continuous contributions can be potentially constrained. We compute the sensitivity (at $2 \sigma$ level), assuming one year integration time, and following~\cite{S5}. For comparison purpose, we also show the bounds from Big Bang Nucleosynthesis (BBN) and CMB observations, including the future Planck experiment \cite{sie07,S5} on the GW stochastic background formed before BBN ($z \sim 5\times 10^{9}$) and before the CMB photons decoupled ($z \sim 1100$) respectively.
As we can see there is a large area where it could be possible to constrain both signatures. 
For low values of $G\mu$ (left side in the $\varepsilon$-$G\mu$ plane) the individual 
SNR contributions (popcorn and continuous) are small and are not expected to be detected. For large $G\mu$  and small $\varepsilon$ values (bottom right region) 
both popcorn and continuous contributions are likely to be probed. 

\begin{figure}
\centering
\includegraphics[angle=0,width=0.45\columnwidth]{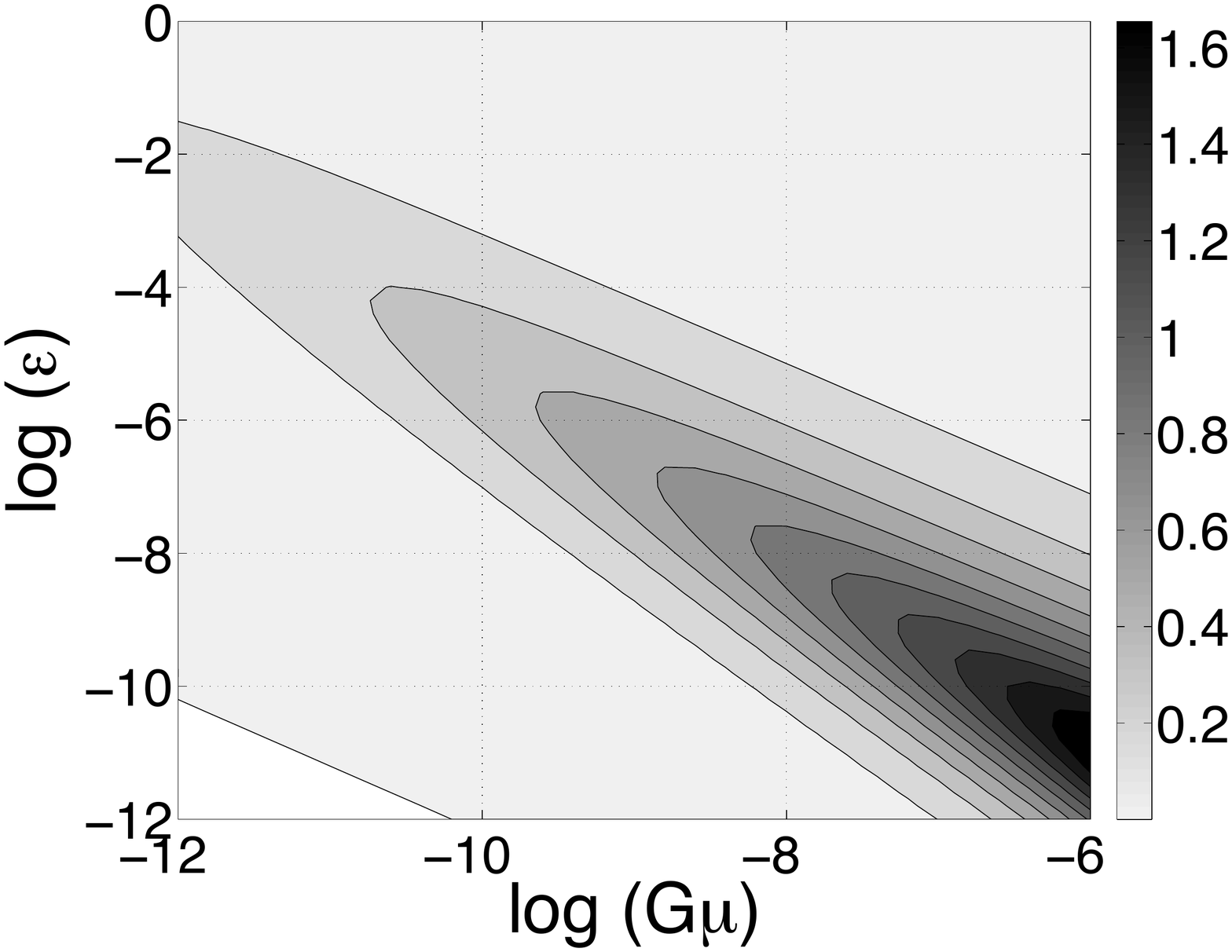}
\includegraphics[angle=0,width=0.45\columnwidth]{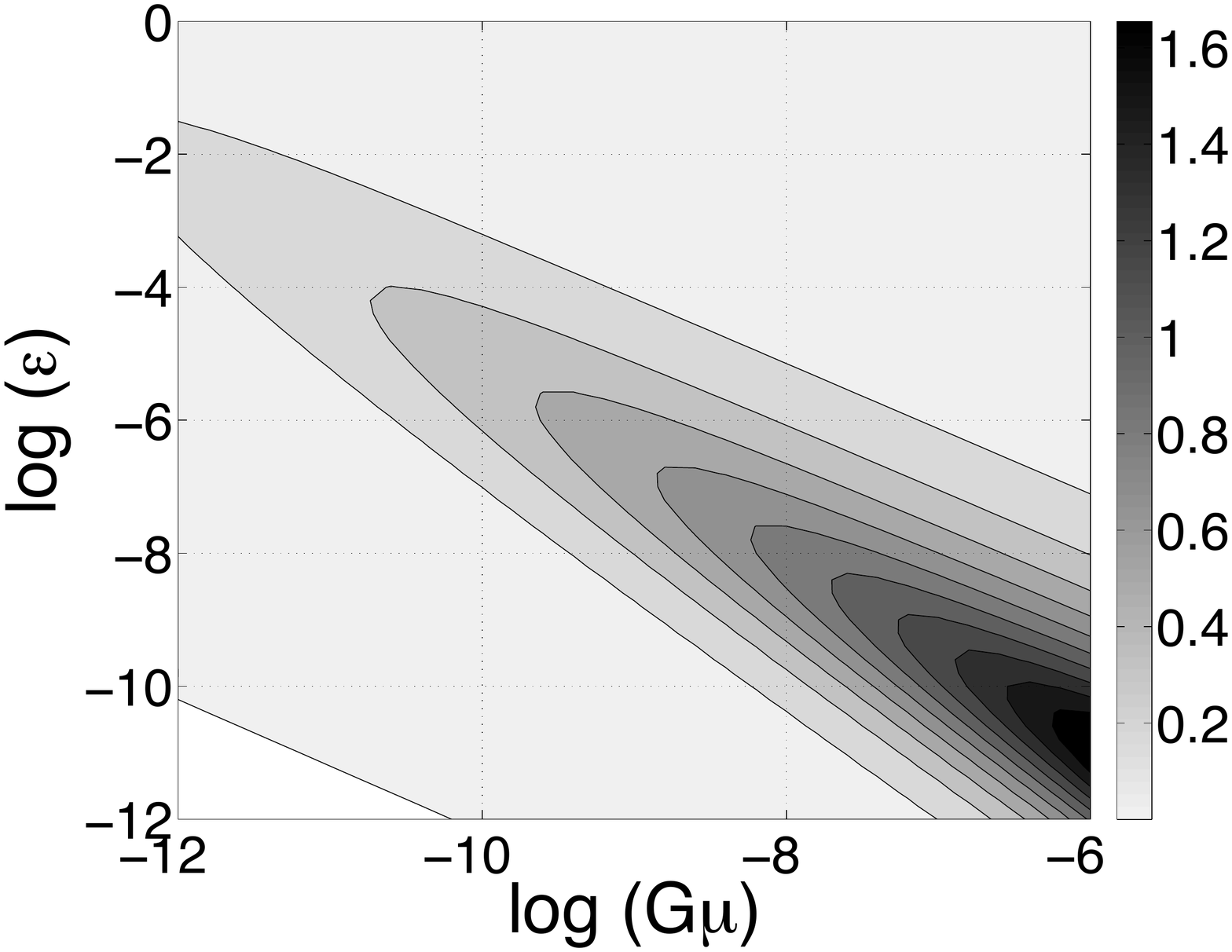}\\
\includegraphics[angle=0,width=0.45\columnwidth]{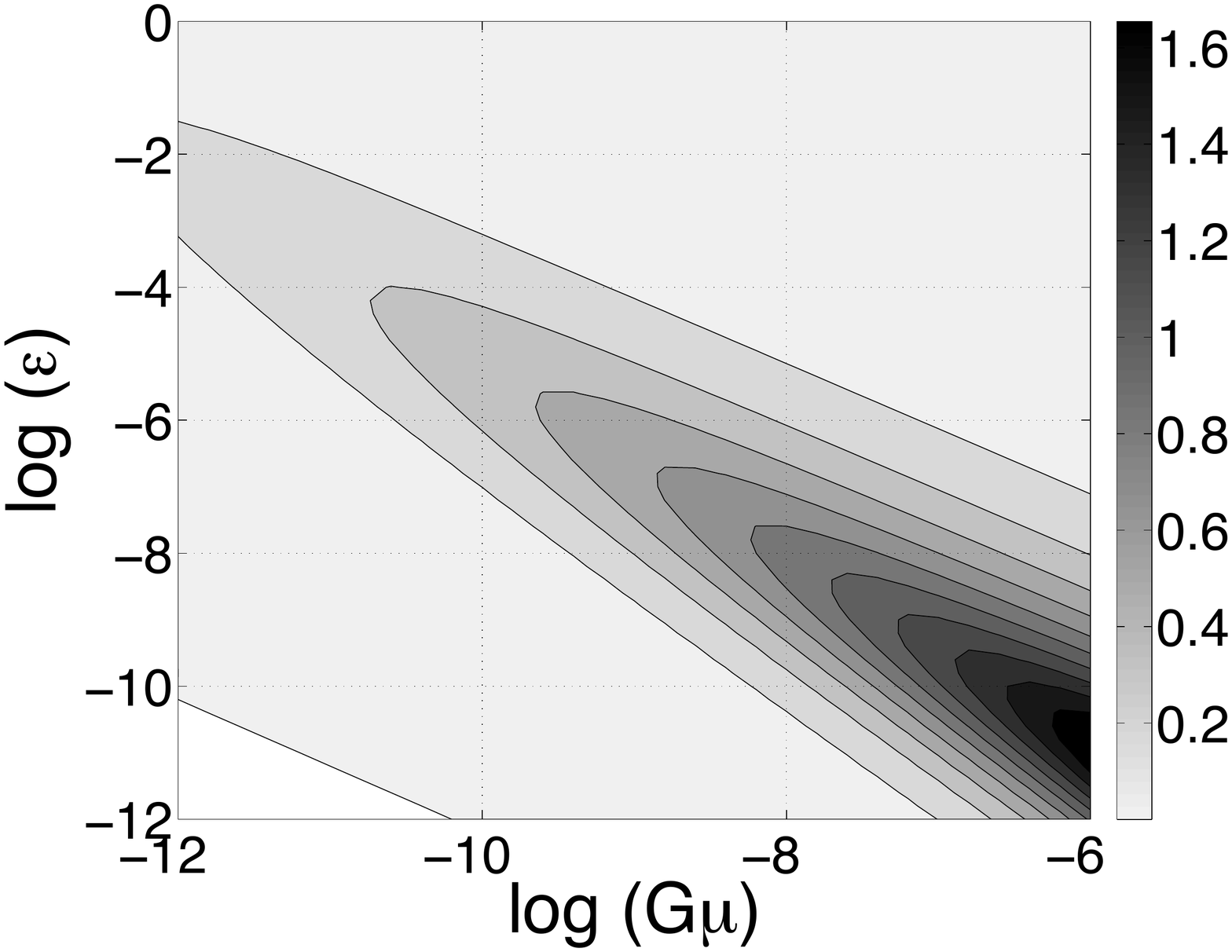}
\includegraphics[angle=0,width=0.45\columnwidth]{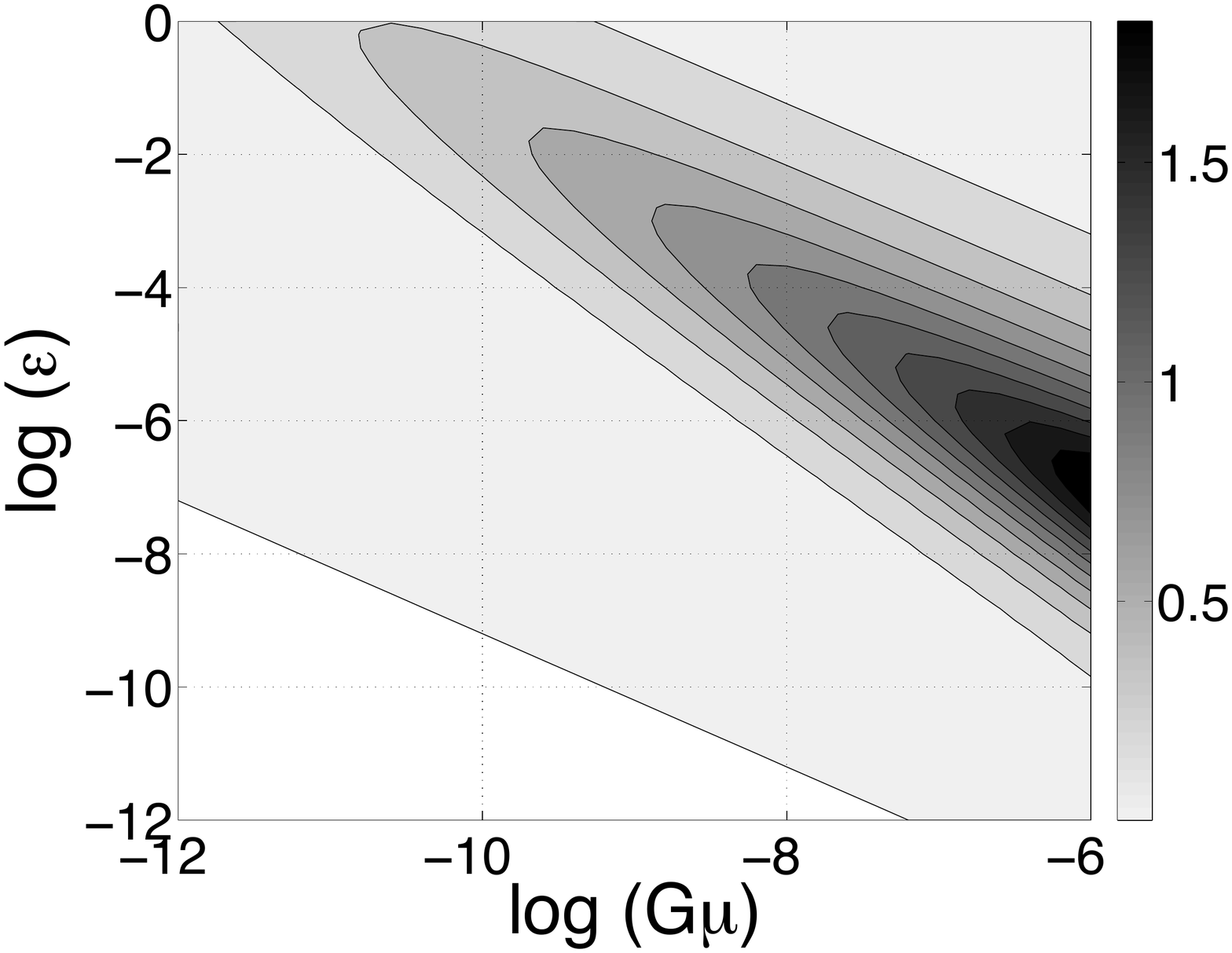}\\
\caption{ratio $R=SNR_{pop} / SNR_{cont}$ for $p=1$, for a pair of separated AdLV (Livingston and Hanford LIGO detectors), a pair of coincident and co-located (CC) AdLV (the 2 LIGO Hanford detectors) (top right), two V-shaped co-located ET detectors (bottom left) and LISA (bottom right). In the denser black regions the popcorn background has a larger contribution in the signal-to-noise ratio. 
In the lighter grey regions, on the other hand, the continuous contribution dominates.
In the white region in the bottom left corner, there is negligible GW signal from cosmic strings in the considered frequency range. \label{fig:snr}}
\end{figure}

Figure~\ref{fig:snr} shows contours in the plane $\varepsilon$-$G\mu$ with constant ratio between the popcorn and continuous contributions to the SNR, for ground based detectors and for LISA. For high values of $p$, the popcorn contribution can 
overwhelm the continuous background in some region of the parameter space.
Even though there is a non negligible region where the popcorn background gives the largest SNR (see Fig ~\ref{fig:snr}), the area 
of the parameter space probed by the popcorn contribution is always included in the area probed by the continuous background. This may change in the future with the 
development of specific data analysis techniques that could perform up to a few times better than the standard cross correlation statistic 
used in this paper \cite{dra03, set09,cow05}. 
Also, for cosmologies other than the standard one used in this paper, for example (see Appendix A), if the redshift at the transition between the matter and radiation era 
$z_{eq}$ were larger than $\sim 4000$, there would be a small region where terrestrial detectors could detect the popcorn contribution only. 

%

\section{Conclusion}
The GW emission from the population of cusps of small loops of cosmic strings fall into two regimes with very different statistical properties: sources at large redshift overlap to create a Gaussian and continuous stochastic background, while close sources create a non-Gaussian and non-continuous popcorn-like signal. The Gaussian continuous background is completely characterized by its spectral properties and can be detected by the standard cross correlation methods in the frequency domain. The popcorn background is less predictable, as it may show important variations in the time domain.

In this paper, we investigated the popcorn and continuous (Gaussian) contribution to the background mapping it onto the cosmic string parameter space. 
The transition between the two regimes depends on the frequency (the larger the frequency, the smaller the transition redshift $z_*$)  and on the cosmic string parameters, in particular on the reconnection probability $p$ (the larger is this parameter, the larger is the rate and the smaller is $z_*$). 

We found that the popcorn contribution may dominate in different regions of the parameter space and over different frequency ranges. It is therefore worthwhile to develop data analyses methods that can better capture the popcorn signature as well as the continuous. Future gravitational wave experiments, such as Advanced LIGO/Virgo, Einstein Telescope, LISA or PTA, may be able to observe both gravitational wave signatures. 

We computed the sensitivity (at $2 \sigma$ level) in the cosmic strings parameter space for AdLV, ET, LISA and PTA. The deduced regions cover a large area of the parameter space but the popcorn contribution to the stochastic background is expected to be more pronounced at higher frequencies where ground based detectors operate. The case of large loop cosmic strings, of relevance to LISA and PTA, will be the subject of a following study. 

In this paper we used the same method for the popcorn and for the continuous backgrounds, but specific data analysis techniques that could perform up to a few times better than the cross-correlation statistics when the background is not Gaussian have been proposed  \cite{dra03, set09,cow05} and are currently investigated in the LIGO/Virgo collaboration. 

An important aspect of detecting the popcorn and continuous background, which further motivates the development of specific data analysis techniques in this case, is the possibility of pinning down the parameter space as additional statistical information is carried by the popcorn sector 
of the spectrum. As shown in this work, in part of the parameter space the relative contribution of the popcorn and continuous backgrounds to the measured signal to noise ratio crucially depends on both the string tension and the reconnection probability. Breaking the degeneracy in these 
parameters requires, as expected, two measurements. Further work on parameter estimation is currently pursued by some of the authors.

\begin{acknowledgments}
The authors thank J.Romano and G. Cella for careful reading and valuable comments. 
X.S.\ acknowledges the support from NSF grant PHY-0970074, PHY-0955929 and PHY-0758155. 
S. G.\  acknowledges the support from NSF grant PHY-0970074 and UWM's Research Growth Initiative.
\end{acknowledgments}

\appendix
\section{Cosmological functions}

The dimensionless cosmological functions $\varphi_t(z)$,  $\varphi_r(z)$ and
$\varphi_V(z)$ were calculated using a vanilla $\Lambda$-CDM model.
We adopted the cosmological parameters derived from 7 years of WMAP observations~\cite{lar11}:  
$H_0=72$  km s$^{-1}$ for the Hubble parameter,  $\Omega_{m}=0.279$ for the density of matter, $\Omega_{r}= 8.5\times10^{-5}$ for the energy density of radiation, and assuming a flat universe $\Omega_{\Lambda}=1-\Omega_{m}-\Omega_{r}$.

The dimensionless cosmological time at redshift $z$ is given by:
\begin{equation}
\varphi_t(z) = \int_0^{z} \frac{dz'}{(1 + z')E(\Omega, z')}\;,
\label{eq:phit}
\end{equation}
where
\begin{equation}
E(\Omega,z)=\sqrt{\Omega_{\Lambda}+\Omega_{m}(1+z)^3+\Omega_{r}(1+z)^4}
\label{eq-E-z}
\end{equation}

The dimensionless proper distance at redshift $z$ by:
\begin{equation}
\varphi_r(z) = \int_0^z \frac{dz'}{E(\Omega, z')}
\label{eq:phir}
\end{equation}

And the dimensionless volume at redshift $z$ by:
\begin{equation}
\varphi_V(z)=4 \pi \frac{\varphi_r(z)^2}{(1+z)^3E(\Omega,z)}
\label{eq:phiV}
\end{equation}

The redshift at the transition between the matter and the radiation dominated eras can be deduced form Eq. ~\ref{eq-E-z}:
\begin{equation}
z_{eq}=\frac{\Omega_{m}}{\Omega_{r}}-1 \sim 3400
\end{equation}

\bibliography{CSbib}

\begin{thebibliography}{36}
\expandafter\ifx\csname natexlab\endcsname\relax\def\natexlab#1{#1}\fi
\expandafter\ifx\csname bibnamefont\endcsname\relax
  \def\bibnamefont#1{#1}\fi
\expandafter\ifx\csname bibfnamefont\endcsname\relax
  \def\bibfnamefont#1{#1}\fi
\expandafter\ifx\csname citenamefont\endcsname\relax
  \def\citenamefont#1{#1}\fi
\expandafter\ifx\csname url\endcsname\relax
  \def\url#1{\texttt{#1}}\fi
\expandafter\ifx\csname urlprefix\endcsname\relax\def\urlprefix{URL }\fi
\providecommand{\bibinfo}[2]{#2}
\providecommand{\eprint}[2][]{\url{#2}}

\bibitem[{\citenamefont{Vilenkin and Shellard}(2000)}]{alexbook}
\bibinfo{author}{\bibfnamefont{A.}~\bibnamefont{Vilenkin}} \bibnamefont{and}
  \bibinfo{author}{\bibfnamefont{E.}~\bibnamefont{Shellard}},
  \emph{\bibinfo{title}{Cosmic strings and other Topological Defects}}
  (\bibinfo{publisher}{Cambridge University Press}, \bibinfo{year}{2000}).

\bibitem[{\citenamefont{Hindmarsh and Kibble}(1995)}]{hin95}
\bibinfo{author}{\bibfnamefont{M.~B.} \bibnamefont{Hindmarsh}}
  \bibnamefont{and} \bibinfo{author}{\bibfnamefont{T.~W.~B.}
  \bibnamefont{Kibble}}, \bibinfo{journal}{Reports on Progress in Physics}
  \textbf{\bibinfo{volume}{58}}, \bibinfo{pages}{477} (\bibinfo{year}{1995}),
  \urlprefix\url{http://stacks.iop.org/0034-4885/58/i=5/a=001}.

\bibitem[{\citenamefont{{Jones} et~al.}(2002)\citenamefont{{Jones}, {Stoica},
  and {Tye}}}]{jon02}
\bibinfo{author}{\bibfnamefont{N.}~\bibnamefont{{Jones}}},
  \bibinfo{author}{\bibfnamefont{H.}~\bibnamefont{{Stoica}}}, \bibnamefont{and}
  \bibinfo{author}{\bibfnamefont{S.-H.~H.} \bibnamefont{{Tye}}},
  \bibinfo{journal}{Journal of High Energy Physics}
  \textbf{\bibinfo{volume}{7}}, \bibinfo{pages}{51} (\bibinfo{year}{2002}),
  \eprint{arXiv:hep-th/0203163}.

\bibitem[{\citenamefont{{Jones} et~al.}(2003)\citenamefont{{Jones}, {Stoica},
  and {Tye}}}]{jon03}
\bibinfo{author}{\bibfnamefont{N.~T.} \bibnamefont{{Jones}}},
  \bibinfo{author}{\bibfnamefont{H.}~\bibnamefont{{Stoica}}}, \bibnamefont{and}
  \bibinfo{author}{\bibfnamefont{S.-H.~H.} \bibnamefont{{Tye}}},
  \bibinfo{journal}{Physics Letters B} \textbf{\bibinfo{volume}{563}},
  \bibinfo{pages}{6} (\bibinfo{year}{2003}), \eprint{arXiv:hep-th/0303269}.

\bibitem[{\citenamefont{{Sarangi} and {Tye}}(2002)}]{sar02}
\bibinfo{author}{\bibfnamefont{S.}~\bibnamefont{{Sarangi}}} \bibnamefont{and}
  \bibinfo{author}{\bibfnamefont{S.-H.~H.} \bibnamefont{{Tye}}},
  \bibinfo{journal}{Physics Letters B} \textbf{\bibinfo{volume}{536}},
  \bibinfo{pages}{185} (\bibinfo{year}{2002}), \eprint{arXiv:hep-th/0204074}.

\bibitem[{\citenamefont{{Dvali} and {Vilenkin}}(2004)}]{dva04}
\bibinfo{author}{\bibfnamefont{G.}~\bibnamefont{{Dvali}}} \bibnamefont{and}
  \bibinfo{author}{\bibfnamefont{A.}~\bibnamefont{{Vilenkin}}},
  \bibinfo{journal}{Journal of Cosmology and Astroparticle Physics}
  \textbf{\bibinfo{volume}{3}}, \bibinfo{pages}{10} (\bibinfo{year}{2004}),
  \eprint{arXiv:hep-th/0312007}.

\bibitem[{\citenamefont{{Copeland} et~al.}(2004)\citenamefont{{Copeland},
  {Myers}, and {Polchinski}}}]{cop04}
\bibinfo{author}{\bibfnamefont{E.~J.} \bibnamefont{{Copeland}}},
  \bibinfo{author}{\bibfnamefont{R.~C.} \bibnamefont{{Myers}}},
  \bibnamefont{and}
  \bibinfo{author}{\bibfnamefont{J.}~\bibnamefont{{Polchinski}}},
  \bibinfo{journal}{Journal of High Energy Physics}
  \textbf{\bibinfo{volume}{6}}, \bibinfo{pages}{13} (\bibinfo{year}{2004}),
  \eprint{arXiv:hep-th/0312067}.

\bibitem[{\citenamefont{Damour and Vilenkin}(2000)}]{dam00}
\bibinfo{author}{\bibfnamefont{T.}~\bibnamefont{Damour}} \bibnamefont{and}
  \bibinfo{author}{\bibfnamefont{A.}~\bibnamefont{Vilenkin}},
  \bibinfo{journal}{Phys. Rev. Lett.} \textbf{\bibinfo{volume}{85}},
  \bibinfo{pages}{3761} (\bibinfo{year}{2000}),
  \urlprefix\url{http://link.aps.org/doi/10.1103/PhysRevLett.85.3761}.

\bibitem[{\citenamefont{Damour and Vilenkin}(2001)}]{dam01}
\bibinfo{author}{\bibfnamefont{T.}~\bibnamefont{Damour}} \bibnamefont{and}
  \bibinfo{author}{\bibfnamefont{A.}~\bibnamefont{Vilenkin}},
  \bibinfo{journal}{Phys. Rev. D} \textbf{\bibinfo{volume}{64}},
  \bibinfo{pages}{064008} (\bibinfo{year}{2001}),
  \urlprefix\url{http://link.aps.org/doi/10.1103/PhysRevD.64.064008}.

\bibitem[{\citenamefont{Damour and Vilenkin}(2005)}]{dam05}
\bibinfo{author}{\bibfnamefont{T.}~\bibnamefont{Damour}} \bibnamefont{and}
  \bibinfo{author}{\bibfnamefont{A.}~\bibnamefont{Vilenkin}},
  \bibinfo{journal}{Phys. Rev. D} \textbf{\bibinfo{volume}{71}},
  \bibinfo{pages}{063510} (\bibinfo{year}{2005}),
  \urlprefix\url{http://link.aps.org/doi/10.1103/PhysRevD.71.063510}.

\bibitem[{\citenamefont{Siemens et~al.}(2006)\citenamefont{Siemens, Creighton,
  Maor, Majumder, Cannon, and Read}}]{sie06}
\bibinfo{author}{\bibfnamefont{X.}~\bibnamefont{Siemens}},
  \bibinfo{author}{\bibfnamefont{J.}~\bibnamefont{Creighton}},
  \bibinfo{author}{\bibfnamefont{I.}~\bibnamefont{Maor}},
  \bibinfo{author}{\bibfnamefont{S.~R.} \bibnamefont{Majumder}},
  \bibinfo{author}{\bibfnamefont{K.}~\bibnamefont{Cannon}}, \bibnamefont{and}
  \bibinfo{author}{\bibfnamefont{J.}~\bibnamefont{Read}},
  \bibinfo{journal}{Phys. Rev. D} \textbf{\bibinfo{volume}{73}},
  \bibinfo{pages}{105001} (\bibinfo{year}{2006}),
  \urlprefix\url{http://link.aps.org/doi/10.1103/PhysRevD.73.105001}.

\bibitem[{\citenamefont{Siemens et~al.}(2007)\citenamefont{Siemens, Mandic, and
  Creighton}}]{sie07}
\bibinfo{author}{\bibfnamefont{X.}~\bibnamefont{Siemens}},
  \bibinfo{author}{\bibfnamefont{V.}~\bibnamefont{Mandic}}, \bibnamefont{and}
  \bibinfo{author}{\bibfnamefont{J.}~\bibnamefont{Creighton}},
  \bibinfo{journal}{Phys. Rev. Lett.} \textbf{\bibinfo{volume}{98}},
  \bibinfo{pages}{111101} (\bibinfo{year}{2007}),
  \urlprefix\url{http://link.aps.org/doi/10.1103/PhysRevLett.98.111101}.

\bibitem[{\citenamefont{Harry and the LIGO
  Scientific~Collaboration}(2010)}]{har10}
\bibinfo{author}{\bibfnamefont{G.~M.} \bibnamefont{Harry}} \bibnamefont{and}
  \bibinfo{author}{\bibnamefont{the LIGO Scientific~Collaboration}},
  \bibinfo{journal}{Classical and Quantum Gravity}
  \textbf{\bibinfo{volume}{27}}, \bibinfo{pages}{084006}
  (\bibinfo{year}{2010}).

\bibitem[{\citenamefont{the Advanced LIGO~Team}(2007)}]{AdLIGO}
\bibinfo{author}{\bibnamefont{the Advanced LIGO~Team}} (\bibinfo{year}{2007}),
  \urlprefix\url{https://dcc.ligo.org/cgi-bin/DocDB/ShowDocument?docid=m060056}.

\bibitem[{\citenamefont{Losurdo and the Advanced Virgo~Team}(2007)}]{AdVIRGO}
\bibinfo{author}{\bibfnamefont{G.}~\bibnamefont{Losurdo}} \bibnamefont{and}
  \bibinfo{author}{\bibnamefont{the Advanced Virgo~Team}}
  (\bibinfo{year}{2007}), \urlprefix\url{https://tds.ego-gw.it/ql/?c=1900}.

\bibitem[{\citenamefont{Punturo et~al.}(2010)}]{ET}
\bibinfo{author}{\bibfnamefont{M.}~\bibnamefont{Punturo}} \bibnamefont{et~al.},
  \bibinfo{journal}{Classical and Quantum Gravity}
  \textbf{\bibinfo{volume}{27}}, \bibinfo{pages}{194002}
  (\bibinfo{year}{2010}),
  \urlprefix\url{http://stacks.iop.org/0264-9381/27/i=19/a=194002}.

\bibitem[{\citenamefont{Bender and the LISA Study~Team}(1998)}]{LISA}
\bibinfo{author}{\bibfnamefont{P.~L.} \bibnamefont{Bender}} \bibnamefont{and}
  \bibinfo{author}{\bibnamefont{the LISA Study~Team}} (\bibinfo{year}{1998}),
  \urlprefix\url{http://list.caltech.edu/lib/exe/fetch.php?media=documents:early:prephasea.pdf}.

\bibitem[{\citenamefont{Manchester}(2011)}]{PTA}
\bibinfo{author}{\bibfnamefont{R.~N.} \bibnamefont{Manchester}}, in
  \emph{\bibinfo{booktitle}{American Institute of Physics Conference Series}},
  edited by \bibinfo{editor}{\bibfnamefont{M.~B.} \bibnamefont{et~al.}}
  (\bibinfo{year}{2011}), vol. \bibinfo{volume}{1357}, pp.
  \bibinfo{pages}{65--72}, \eprint{1101.5202}.

\bibitem[{\citenamefont{{McGraw}}(1998)}]{1998PhRvD..57.3317M}
\bibinfo{author}{\bibfnamefont{P.}~\bibnamefont{{McGraw}}},
  \bibinfo{journal}{\prd} \textbf{\bibinfo{volume}{57}}, \bibinfo{pages}{3317}
  (\bibinfo{year}{1998}), \eprint{arXiv:astro-ph/9706182}.

\bibitem[{\citenamefont{{Spergel} and {Pen}}(1997)}]{1997ApJ...491L..67S}
\bibinfo{author}{\bibfnamefont{D.}~\bibnamefont{{Spergel}}} \bibnamefont{and}
  \bibinfo{author}{\bibfnamefont{U.-L.} \bibnamefont{{Pen}}},
  \bibinfo{journal}{ApJL} \textbf{\bibinfo{volume}{491}}, \bibinfo{pages}{L67}
  (\bibinfo{year}{1997}), \eprint{arXiv:astro-ph/9611198}.

\bibitem[{\citenamefont{Dubath et~al.}(2008)\citenamefont{Dubath, Polchinski,
  and Rocha}}]{Dubath:2007mf}
\bibinfo{author}{\bibfnamefont{F.}~\bibnamefont{Dubath}},
  \bibinfo{author}{\bibfnamefont{J.}~\bibnamefont{Polchinski}},
  \bibnamefont{and} \bibinfo{author}{\bibfnamefont{J.~V.} \bibnamefont{Rocha}},
  \bibinfo{journal}{Phys.Rev.} \textbf{\bibinfo{volume}{D77}},
  \bibinfo{pages}{123528} (\bibinfo{year}{2008}), \eprint{0711.0994}.

\bibitem[{\citenamefont{Siemens et~al.}(2002)\citenamefont{Siemens, Olum, and
  Vilenkin}}]{Siemens:2002dj}
\bibinfo{author}{\bibfnamefont{X.}~\bibnamefont{Siemens}},
  \bibinfo{author}{\bibfnamefont{K.~D.} \bibnamefont{Olum}}, \bibnamefont{and}
  \bibinfo{author}{\bibfnamefont{A.}~\bibnamefont{Vilenkin}},
  \bibinfo{journal}{Phys.Rev.} \textbf{\bibinfo{volume}{D66}},
  \bibinfo{pages}{043501} (\bibinfo{year}{2002}), \eprint{gr-qc/0203006}.

\bibitem[{\citenamefont{Blanco-Pillado
  et~al.}(2011)\citenamefont{Blanco-Pillado, Olum, and
  Shlaer}}]{BlancoPillado:2011dq}
\bibinfo{author}{\bibfnamefont{J.~J.} \bibnamefont{Blanco-Pillado}},
  \bibinfo{author}{\bibfnamefont{K.~D.} \bibnamefont{Olum}}, \bibnamefont{and}
  \bibinfo{author}{\bibfnamefont{B.}~\bibnamefont{Shlaer}},
  \bibinfo{journal}{Phys.Rev.} \textbf{\bibinfo{volume}{D83}},
  \bibinfo{pages}{083514} (\bibinfo{year}{2011}), \eprint{1101.5173}.

\bibitem[{\citenamefont{Allen and Romano}(1999)}]{all99}
\bibinfo{author}{\bibfnamefont{B.}~\bibnamefont{Allen}} \bibnamefont{and}
  \bibinfo{author}{\bibfnamefont{J.~D.} \bibnamefont{Romano}},
  \bibinfo{journal}{Phys. Rev. D} \textbf{\bibinfo{volume}{59}},
  \bibinfo{pages}{102001} (\bibinfo{year}{1999}),
  \urlprefix\url{http://link.aps.org/doi/10.1103/PhysRevD.59.102001}.

\bibitem[{\citenamefont{{Turok}}(1984)}]{1984NuPhB.242..520T}
\bibinfo{author}{\bibfnamefont{N.}~\bibnamefont{{Turok}}},
  \bibinfo{journal}{Nuclear Physics B} \textbf{\bibinfo{volume}{242}},
  \bibinfo{pages}{520} (\bibinfo{year}{1984}).

\bibitem[{\citenamefont{Buonanno et~al.}(2005)\citenamefont{Buonanno, Sigl,
  Raffelt, Janka, and M\"uller}}]{buo05}
\bibinfo{author}{\bibfnamefont{A.}~\bibnamefont{Buonanno}},
  \bibinfo{author}{\bibfnamefont{G.}~\bibnamefont{Sigl}},
  \bibinfo{author}{\bibfnamefont{G.~G.} \bibnamefont{Raffelt}},
  \bibinfo{author}{\bibfnamefont{H.-T.} \bibnamefont{Janka}}, \bibnamefont{and}
  \bibinfo{author}{\bibfnamefont{E.}~\bibnamefont{M\"uller}},
  \bibinfo{journal}{Phys. Rev. D} \textbf{\bibinfo{volume}{72}},
  \bibinfo{pages}{084001} (\bibinfo{year}{2005}),
  \urlprefix\url{http://link.aps.org/doi/10.1103/PhysRevD.72.084001}.

\bibitem[{\citenamefont{{Rosado}}(2011)}]{ros11}
\bibinfo{author}{\bibfnamefont{P.~A.} \bibnamefont{{Rosado}}},
  \bibinfo{journal}{\prd} \textbf{\bibinfo{volume}{84}}, \bibinfo{eid}{084004}
  (\bibinfo{year}{2011}), \eprint{1106.5795}.

\bibitem[{\citenamefont{Coward and Regimbau}(2006)}]{cow06}
\bibinfo{author}{\bibfnamefont{D.}~\bibnamefont{Coward}} \bibnamefont{and}
  \bibinfo{author}{\bibfnamefont{T.}~\bibnamefont{Regimbau}},
  \bibinfo{journal}{New Astronomy Reviews} p. \bibinfo{pages}{461}
  (\bibinfo{year}{2006}).

\bibitem[{\citenamefont{Drasco and Flanagan}(2003)}]{dra03}
\bibinfo{author}{\bibfnamefont{S.}~\bibnamefont{Drasco}} \bibnamefont{and}
  \bibinfo{author}{\bibfnamefont{E.~E.} \bibnamefont{Flanagan}},
  \bibinfo{journal}{Phys. Rev. D} \textbf{\bibinfo{volume}{67}},
  \bibinfo{pages}{082003} (\bibinfo{year}{2003}),
  \urlprefix\url{http://link.aps.org/doi/10.1103/PhysRevD.67.082003}.

\bibitem[{\citenamefont{Jenet et~al.}(2006)\citenamefont{Jenet, Hobbs, van
  Straten, Manchester, Bailes, Verbiest, Edwards, Hotan, Sarkissian, and
  Ord}}]{Jenet06}
\bibinfo{author}{\bibfnamefont{F.~A.} \bibnamefont{Jenet}},
  \bibinfo{author}{\bibfnamefont{G.~B.} \bibnamefont{Hobbs}},
  \bibinfo{author}{\bibfnamefont{W.}~\bibnamefont{van Straten}},
  \bibinfo{author}{\bibfnamefont{R.~N.} \bibnamefont{Manchester}},
  \bibinfo{author}{\bibfnamefont{M.}~\bibnamefont{Bailes}},
  \bibinfo{author}{\bibfnamefont{J.~P.~W.} \bibnamefont{Verbiest}},
  \bibinfo{author}{\bibfnamefont{R.~T.} \bibnamefont{Edwards}},
  \bibinfo{author}{\bibfnamefont{A.~W.} \bibnamefont{Hotan}},
  \bibinfo{author}{\bibfnamefont{J.~M.} \bibnamefont{Sarkissian}},
  \bibnamefont{and} \bibinfo{author}{\bibfnamefont{S.~M.} \bibnamefont{Ord}},
  \bibinfo{journal}{The Astrophysical Journal} \textbf{\bibinfo{volume}{653}},
  \bibinfo{pages}{1571} (\bibinfo{year}{2006}),
  \urlprefix\url{http://stacks.iop.org/0004-637X/653/i=2/a=1571}.

\bibitem[{\citenamefont{{Hogan} and {Bender}}(2001)}]{hogan01}
\bibinfo{author}{\bibfnamefont{C.~J.} \bibnamefont{{Hogan}}} \bibnamefont{and}
  \bibinfo{author}{\bibfnamefont{P.~L.} \bibnamefont{{Bender}}},
  \bibinfo{journal}{Phys. Rev. D} \textbf{\bibinfo{volume}{64}},
  \bibinfo{pages}{062002} (\bibinfo{year}{2001}).

\bibitem[{\citenamefont{Vecchio}(2002)}]{vec02}
\bibinfo{author}{\bibfnamefont{A.}~\bibnamefont{Vecchio}},
  \bibinfo{journal}{Classical and Quantum Gravity}
  \textbf{\bibinfo{volume}{19}}, \bibinfo{pages}{1449} (\bibinfo{year}{2002}),
  \urlprefix\url{http://stacks.iop.org/0264-9381/19/i=7/a=329}.

\bibitem[{\citenamefont{Abbott et~al.}(2009)}]{S5}
\bibinfo{author}{\bibfnamefont{B.~P.} \bibnamefont{Abbott}}
  \bibnamefont{et~al.}, \bibinfo{journal}{Nature}
  \textbf{\bibinfo{volume}{460}}, \bibinfo{pages}{990} (\bibinfo{year}{2009}),
  \urlprefix\url{http://dx.doi.org/10.1038/nature08278}.

\bibitem[{\citenamefont{Seto}(2009)}]{set09}
\bibinfo{author}{\bibfnamefont{N.}~\bibnamefont{Seto}}, \bibinfo{journal}{Phys.
  Rev. D} \textbf{\bibinfo{volume}{80}}, \bibinfo{pages}{043003}
  (\bibinfo{year}{2009}),
  \urlprefix\url{http://link.aps.org/doi/10.1103/PhysRevD.80.043003}.

\bibitem[{\citenamefont{Coward and Burman}(2005)}]{cow05}
\bibinfo{author}{\bibfnamefont{D.~M.} \bibnamefont{Coward}} \bibnamefont{and}
  \bibinfo{author}{\bibfnamefont{R.~R.} \bibnamefont{Burman}},
  \bibinfo{journal}{Monthly Notices of the Royal Astronomical Society}
  \textbf{\bibinfo{volume}{361}}, \bibinfo{pages}{362} (\bibinfo{year}{2005}),
  ISSN \bibinfo{issn}{1365-2966},
  \urlprefix\url{http://dx.doi.org/10.1111/j.1365-2966.2005.09178.x}.

\bibitem[{\citenamefont{{Larson} et~al.}(2011)\citenamefont{{Larson},
  {Dunkley}, {Hinshaw}, {Komatsu}, {Nolta}, {Bennett}, {Gold}, {Halpern},
  {Hill}, {Jarosik} et~al.}}]{lar11}
\bibinfo{author}{\bibfnamefont{D.}~\bibnamefont{{Larson}}},
  \bibinfo{author}{\bibfnamefont{J.}~\bibnamefont{{Dunkley}}},
  \bibinfo{author}{\bibfnamefont{G.}~\bibnamefont{{Hinshaw}}},
  \bibinfo{author}{\bibfnamefont{E.}~\bibnamefont{{Komatsu}}},
  \bibinfo{author}{\bibfnamefont{M.~R.} \bibnamefont{{Nolta}}},
  \bibinfo{author}{\bibfnamefont{C.~L.} \bibnamefont{{Bennett}}},
  \bibinfo{author}{\bibfnamefont{B.}~\bibnamefont{{Gold}}},
  \bibinfo{author}{\bibfnamefont{M.}~\bibnamefont{{Halpern}}},
  \bibinfo{author}{\bibfnamefont{R.~S.} \bibnamefont{{Hill}}},
  \bibinfo{author}{\bibfnamefont{N.}~\bibnamefont{{Jarosik}}},
  \bibnamefont{et~al.}, \bibinfo{journal}{Astrophysical Journal Supplement}
  \textbf{\bibinfo{volume}{192}}, \bibinfo{pages}{16} (\bibinfo{year}{2011}),
  \eprint{1001.4635}.

\end{thebibliography}

\end{document}